\documentclass[11pt,a4paper]{article}
\usepackage{graphicx}
\usepackage{jheppub}
\usepackage{setspace}
\usepackage{amsmath,amssymb,color,mathrsfs,verbatim,bbm,wasysym,pstricks,epsfig,slashed}

\title{New light species and the CMB}
\author[a,b]{Christopher Brust,}
\author[a]{David E.\ Kaplan,}
\author[a]{and Matthew T.\ Walters}
\affiliation[a]{Department of Physics and Astronomy, Johns Hopkins University, \\
Charles Street, Baltimore, MD 21218, U.S.A.}
\affiliation[b]{Department of Physics, University of Maryland, \\
Campus Drive, College Park, MD 20742, U.S.A.}
\emailAdd{cbrust@pha.jhu.edu}
\emailAdd{dkaplan@pha.jhu.edu}
\emailAdd{mwalters@pha.jhu.edu}
\abstract{We consider the effects of new light species on the Cosmic Microwave Background. In the massless limit, these effects can be parameterized in terms of a single number, the relativistic degrees of freedom. We perform a thorough survey of natural, minimal models containing new light species and numerically calculate the precise contribution of each of these models to this number in the framework of effective field theory. After reviewing the relevant details of early universe thermodynamics, we provide a map between the parameters of any particular theory and the predicted effective number of degrees of freedom. We then use this map to interpret the recent results from the Cosmic Microwave Background survey done by the Planck satellite. Using this data, we present new constraints on the parameter space of several models containing new light species. Future measurements of the Cosmic Microwave Background can be used with this map to further constrain the parameter space of all such models.}
\arxivnumber{1303.5379}

\begin{document}

\begin{flushright}UMD-PP-013-005\end{flushright}

\maketitle
\flushbottom

\section{Introduction}
\label{sec:intro}

The Cosmic Microwave Background (CMB) is one of the only probes we have of physics in the early universe. Through a detailed mapping of anisotropies in the temperature of those photons which decoupled from visible matter in the era of recombination, we are able to determine the relativistic energy density in that era. From this, we gain information about the number of light species in our universe. In the massless limit, we can accomplish this by a fit to only one number, the relativistic degrees of freedom $g_*$ \cite{JungmanEtAl,Dolgov,Steigman01,LopezEtAl,RiemerSorensonEtAl}. This parameter is often expressed in terms of an effective number of neutrinos, $N_{eff}$, defined such that in the Standard Model (SM) of particle physics $N_{eff}$ is roughly the number of neutrino generations. Beyond the Standard Model (BSM) physics models which contain new light species with masses $\mathcal{O}($eV$)$ or less can contribute to this measurement. Consequently, we have new terrain in which to test the SM through its prediction of $g_* = 3.38$, corresponding to an $N_{eff}$ of $3.046$ \cite{GnedinEtAl,ManganoEtAl,HannestadMadsen,YuehBuchler,Heckler, DolgovFukugita,DolgovEtAl,DodelsonEtAl,RanaEtAl,HerreraEtAl,EspositoEtAl}.

There has been a statistically insignificant but consistent excess in the measured value of $g_*$ \cite{HinshawEtAl,SieversEtAl,HouEtAl12, DiValentinoEtAl}. Prior to the results from the Planck satellite, the most precise reported measurement was $g_* = 3.69 \pm 0.16$, corresponding to $N_{eff} = 3.71 \pm 0.35$ \cite{HouEtAl12}, coming from a combination of data from the South Pole Telescope (SPT) and the Wilkinson Microwave Anisotropy Probe (WMAP). A similar excess is present in measurements from the Atacama Cosmology Telescope (ACT) \cite{SieversEtAl}. Very recently, however, the Planck collaboration released the first results from its measurement of CMB anisotropies, obtaining a result of $g_* = 3.50 \pm 0.12$, corresponding to $N_{eff} = 3.30 \pm 0.27$ \cite{Planck}. Future Planck results will continue to improve the precision of this measurement, with a projected final $g_*$ sensitivity of $\pm 0.09$ \cite{Hamann:2007sb,GalliEtAl}. In addition, future measurements of the polarization of the CMB are projected to constrain $g_*$ to within $\pm 0.02$, corresponding to constraints on $N_{eff}$ of $\pm 0.044$ \cite{GalliEtAl}. We are entering an era of being able to contrast the SM prediction for $g_*$ with the predictions of BSM physics models containing new light species to an unprecedented precision.

The power of this probe of new physics is that in any BSM theory containing new species with masses $\ll 0.1$ eV which were once in thermal equilibrium with the SM, the effect of these species is contained in a single number, the correction $\Delta g_*$ to the SM prediction for $g_*$. Therefore, a map from the parameters of a BSM model to the number $\Delta g_*$ can be constructed in order to determine the consistency of regions of the parameter space with the measured value of $g_*$. Although useful approximations of such a map exist \cite{NakayamaEtAl,DiamantiEtAl}, we are entering the exciting era of precision cosmology experiments, and consequently it has become imperative to form precise theoretical predictions. The subject of this paper is the precise numerical computation of this map of model parameters to $\Delta g_*$ for a wide variety of natural, minimal BSM theories containing new light or massless species. We approach this problem in a largely model-independent effective field theory framework to fully characterize the effects of all such models.

Although there are other existing constraints on new light species present in the early universe coming from the study of Big Bang Nucleosynthesis (BBN) \cite{Steigman07,PospelovEtAl,IoccoEtAl}, this probe does not have the same resolving power as the Planck satellite. Unlike BBN, Planck and future polarization experiments have the power to probe the actual values of the couplings of new light species to the SM, as we shall demonstrate in this work. Even in the absence of a signal for new physics from future experiments, the results of this work provide new constraints on the couplings of SM species to new light particles which are competitive with, and sometimes even surpass, existing constraints from other areas of physics. This establishes a new arena for testing the predictions of BSM physics models with new light species.

The recent results for $g_*$ from Planck are in tension with independent measurements of the Hubble expansion rate today \cite{Planck}. Specifically, combining those measurements with the results from Planck leads to a preference for higher values of $g_*$ than quoted above. Therefore, these results are not capable of confirming or rejecting the hypothesis of new light degrees of freedom being present in the early universe. Regardless, in order to demonstrate the constraining power of measurements of the CMB, we proceed as if this tension were not present. Motivated by the Planck results given above, we proceed by supposing that values of $g_* \geq 3.74$ ($N_{eff} \geq 3.84$) are excluded at the $95\%$ confidence level. We interpret our results in this framework in order to illustrate how further data could be utilized.

In section \ref{sec:methods} of this paper, we review the relevant details of the determination of $g_*$ using the CMB, as well as details of thermodynamics in an expanding universe, providing a framework for the rest of the paper. In section \ref{sec:models} of this paper, we discuss all BSM physics models compatible with our criteria of naturalness and minimality. Specifically, we discuss the parameters which provide the interaction strength between various fields in the SM and the new light species present in the model. We present the current experimental constraints on each of these scenarios, as well as our findings for the contribution of each new light species to $g_*$ as a function of the parameters in the underlying theory. We also interpret the viable parameter space of each model in terms our aforementioned interpretation of the recent results from the Planck satellite, placing additional constraints on theories using this new CMB data.





\section{Methodology}
\label{sec:methods}

We study the effects of adding new light or massless particles to the SM on the evolution of the universe and the CMB. Specifically, we investigate new particles which at some time in the early universe were in equilibrium with the SM and decouple prior to recombination. Translating between additional fields in the Lagrangian and the measurement of the effective number of light degrees of freedom, $g_*$, requires a detailed analysis of the quasi-thermal evolution of the universe. The effects of new light degrees of freedom depend on both when and how they decouple from the thermal bath. As we shall see, a direct measurement of anisotropies in the CMB then leads to a resultant measure of $g_*$ at recombination.

In this section, we first review how light species predominantly affect the CMB, namely via Silk damping and the early integrated Sachs-Wolfe (ISW) effect. We also review the thermodynamics of the early universe, as well as the effects of decoupling and other non-equilibrium events. We then discuss the range of decoupling temperatures which can significantly impact the CMB. Finally, we briefly review the most important existing constraint on new light degrees of freedom, namely their effect on Big Bang Nucleosynthesis. As this section is predominantly a review, readers familiar with early universe thermodynamics can potentially skip to the summary provided in subsection \ref{sec:summary}.


\subsection{Relativistic species and the CMB}

The early universe was not perfectly homogeneous, but instead had small perturbations in the distribution of energy density, which are currently believed to be seeded by inflation. These regions of under- or overdensity correspond to small perturbations in the metric away from the pure Friedmann-Robertson-Walker form. CMB anisotropies provide a direct measurement of these early universe perturbations, whose distribution and structure are sensitive to the thermodynamic conditions leading up to recombination. The CMB therefore gives us insight on the properties and structure of the universe in its infancy. The measurement of $g_*$ using the CMB is performed through a precise determination of the expansion rate, $H$, in the era of recombination. The relationship between $H$ and $g_*$ arises because the expansion rate is determined solely by the total energy density, $\rho$, and the curvature. Increasing the value of $g_*$ at a fixed temperature leads to a larger overall $\rho$, which then leads to more rapid expansion. Silk damping is sensitive to the value of $H$ leading up to and during recombination, while the early ISW effect is affected by the evolution of $H$ once photons are effectively free-streaming, which lasts from recombination onwards. For more detailed and thorough explanations of these effects than those presented here, consult \cite{BowenEtAl,BashinskyEtAl,HouEtAl11} and references therein.


\subsubsection{Silk damping}

Prior to recombination, protons, electrons, and photons interacted very strongly to form a tightly-coupled plasma. Despite the high frequency of interactions, the mean free path for photons was nonzero, and photons were able to diffuse outward. The rate of photon diffusion grew as the protons and electrons combined into hydrogen, up until the point of last scattering. The overall diffusion scale at the end of recombination is therefore predominantly determined by the mean free path during recombination and the duration of recombination. The diffusion of photons results in a partial thermalization of the baryon-photon plasma, damping any inhomogeneities on scales smaller than the photon diffusion length. This reduction of inhomogeneities below some length scale in turn leads to a damping of temperature anisotropies, commonly called Silk damping \cite{Silk}, above some multipole moment $l_d$. A larger value for $H$ then leads to a decrease in the amount of time available for this diffusion, restricting the damping to smaller angular scales and reducing the magnitude of the damping. An increase in $g_*$ would therefore lead to reduced Silk damping, or equivalently a larger damping moment.

Any map between the predicted diffusion length and the precise value for $l_d$ is sensitive to experimental uncertainty in the angular distance to the last scattering surface. In practice, it is simpler to remove this uncertainty by considering the ratio of $l_d$ to the smaller sound horizon moment $l_s$. This sound horizon arises independently of photon diffusion, due to the spread of inhomogeneities in the baryon-photon plasma. These oscillations propagate at the corresponding speed of sound, setting an acoustic oscillation length scale at recombination. The addition of new light species reduces the time for these inhomogeneities to spread, which increases the value of $l_s$, in addition to the increase in $l_d$. These two processes, photon diffusion and sound wave propagation, have different time dependencies. This difference results in an increase of the ratio $l_s/l_d$ as $H$ grows, leading to damping of more of the acoustic peaks, despite the fact that the overall damping has been reduced.


\subsubsection{Early integrated Sachs-Wolfe effect}

Following recombination, photons propagate freely without scattering but pass through points of matter over- or underdensity. If the gravitational potential of these inhomogeneities is constant in time, there is no net effect on the CMB photons. However, if the gravitational potential has any time-dependence, the photons will experience some net loss or gain in energy as they pass through a single gravitational perturbation and be red- or blueshifted as a result. The alteration of CMB anisotropies due to time-dependent gravitational potentials is the ISW effect \cite{SachsWolfe}.

The evolution of gravitational potentials is determined by the expansion rate, which depends on the overall particle content. In a universe consisting solely of nonrelativistic, pressureless matter, the competing effects of gravitational clustering and universe expansion cancel, such that potentials are time-independent. However, any nonnegligible pressure alters the expansion rate such that the potentials do evolve with time. There are therefore two points in time at which the ISW effect could contribute to the CMB. The first occurs when the universe contains a nonnegligible radiation density, which is the case immediately following recombination. This alteration to the CMB shortly after its formation is commonly referred to as the early ISW effect. The second era corresponds to the point at which the vacuum energy becomes a significant fraction of the total energy density. This second case, which begins near modern times, is the late ISW effect.

Unsurprisingly, new light species increase the radiation energy density following recombination, altering $H$ and enhancing the early ISW effect. Specifically, the presence of additional species causes gravitational potentials to evolve more rapidly, resulting in more substantial red- and blueshifts to CMB photons passing through these evolving potentials. On very small scales, photons will pass through multiple such potentials, and the net effect cancels. However, the potentials rapidly become time-independent, such that photons are unable to pass through multiple large-scale perturbations before this effect ends. An increase to $g_*$ therefore enhances the variance in temperature anisotropies on angular scales corresponding to the largest structures immediately following recombination. The size of the largest structures at this point coincides with the acoustic horizon, such that the early ISW effect leads to an increase in the first acoustic peaks of the CMB. In practice, this effect is measured by comparing the height of the first acoustic peak to that of latter peaks.

\begin{figure}[t]
\centering
\includegraphics[width=15cm]{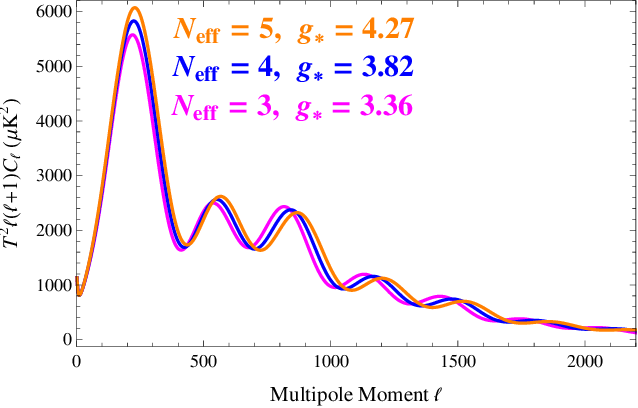}
\caption{Projected CMB anisotropy power spectrum for three different values of $g_*$ (or equivalently $N_{eff}$). The addition of new light degrees of freedom increases the height of the first peak through the early ISW effect and decreases the height of later peaks through Silk damping. The power spectrum, and therefore these effects, are measured by multiple observational experiments, such as the Planck satellite. These spectra were calculated using CAMB \cite{LewisEtAl,HowlettEtAl}. The magenta, blue, and orange curves (dark gray, black, and light gray curves, when viewed in black and white) correspond to an $N_{eff}$ of 3, 4, and 5, respectively.}
\label{fig:CMB}
\end{figure}

The effects of Silk damping and early ISW, which can be seen in figure \ref{fig:CMB}, are complementary means of measuring $H$, and therefore $g_*$, near recombination. However, they are still sensitive to two different points in time. Silk damping probes $H$ prior to and during recombination, while the early ISW effect is sensitive to $H$ immediately after recombination. This has two important consequences for constraints on light species. The first is that experiments which focus on precision measurement at smaller values of $l$, such as WMAP, are sensitive mainly to the early ISW effect. The resulting constraints are therefore more limited experimentally by the effects of cosmic variance. Experiments which instead focus on anisotropies at larger $l$, such as ACT, SPT, and Planck, are predominantly sensitive to Silk damping and are less affected by cosmic variance.

The second consequence to note is that particles with masses near the temperature scale of recombination ($\sim 0.1-1$ eV) will potentially contribute very different signals to these two sets of experiments. The detection of such species would involve a detailed analysis of each individual effect, rather than a simple fit to all the experimental data. While we consider massless particles for the majority of this work, we will return to this possibility later in subsection \ref{sec:massive}.


\subsection{Early universe thermodynamics}
\label{sec:thermo}

As mentioned earlier, CMB measurements of light species are predominantly sensitive to the relativistic energy density, which is characterized by an effective number of relativistic degrees of freedom $g_*$. For more details on the material discussed in this subsection, see \cite{KolbTurner,Dodelson,Weinberg}. We can then define $g_*$ in terms of $\rho_{rel}$, the energy density of all relativistic species, and a reference temperature $T$, which we take to be the photon temperature ($T \equiv T_\gamma$),

\begin{equation}
\label{eq:rhorel}
\rho_{rel} \equiv g_*\frac{\pi^2}{30}T^4 \textrm{.}
\end{equation}

The light species content of the SM, which consists of photons and neutrinos, can then be used to make a prediction for the measured value of $g_*$ at recombination. This prediction can be written in the form

\begin{equation}
g_* = g_\gamma + \frac{7}{8} g_\nu N_{eff} \left( \frac{T_\nu}{T} \right)^4 = 2 + \frac{7}{8} \cdot 2 \cdot N_{eff} \left( \frac{4}{11} \right)^{4/3} \textrm{,}
\end{equation}

{\noindent}where the factor of $\frac{7}{8}$ is due to the effect of Fermi-Dirac statistics on energy density, and $T_\nu$ is the calculated neutrino temperature assuming neutrinos instantaneously decouple from the rest of the SM at $T \sim$ MeV. The parameter $N_{eff}$ is the effective number of neutrino species. This historically defined parameter, which is $3.046$\footnote{This effective number of neutrinos is defined such that if neutrinos truly did decouple instantaneously, $N_{eff}$ would be 3. However, detailed calculations have shown this to not be the case, and the actual energy density of neutrinos is slightly larger than in the instantaneous decoupling approximation due to their interactions with annihilating electrons. This then results in the slightly larger predicted value for $N_{eff}$. For details on these calculations, see \cite{GnedinEtAl,ManganoEtAl,HannestadMadsen, YuehBuchler,Heckler,DolgovFukugita,DolgovEtAl, DodelsonEtAl,RanaEtAl,HerreraEtAl,EspositoEtAl}.} for the SM, is often used to parametrize the effect of any light species other than photons on $g_*$. The contribution of neutrinos and any new light\footnote{By light, we mean $m \ll $ eV. The contribution of species with masses $\sim$ eV is more complicated, as we shall discuss later.} species to $g_*$ is given solely by $N_{eff}$. Any measured deviation from the SM prediction of $g_* = 3.38$ would then indicate the need for new physics.

This paper calculates the full contribution $\Delta g_*$ of new light species present in a large number of beyond the SM theories. This contribution to the relativistic degrees of freedom is found by calculating the energy density of new species near the point of recombination. The contribution can also be expressed as a change to $N_{eff}$ as

\begin{equation}
\Delta N_{eff} = \frac{8}{7} \frac{\Delta g_*}{g_\nu} \left( \frac{T}{T_\nu} \right)^4 \approx 2.2 \, \Delta g_* \textrm{.}
\end{equation}

To find the energy density of a light species at recombination, we must track the evolution of its phase space density $f(t,p)$. This form for the distribution function relies on the assumption that the universe is homogeneous and isotropic\footnote{As discussed earlier, the universe is in fact not perfectly homogeneous or isotropic, and the distribution functions therefore have some spatial and directional dependence. However, these deviations are quite small in magnitude, and any resulting correction to the CMB is below the experimental resolution. Consequently, any inhomogeneities and anisotropies in the distribution functions are negligible for our purposes.}. We must first determine $f$ at high temperatures, when the new species is in equilibrium with the SM, then calculate the changes to $f$ as the universe expands and cools, with various species annihilating or decoupling.

As the universe expands, the evolution of each individual phase space density is controlled by both the rate of expansion $H \equiv \frac{\dot{a}}{a}$, where $a(t)$ is the scale factor for the expanding universe, and the rate of interaction with the other particle species. This dependence is expressed using the Boltzmann equation

\begin{equation}
E \frac{\partial f}{\partial t} -  H p^2 \frac{\partial f}{\partial E}  = C[f] \textrm{,}
\label{eq:boltz}
\end{equation}

{\noindent}where $p=|\vec{p}|$ and the collision functional $C[f]$ accounts for changes to $f$ due to interactions. If we assume that the dominant interactions will consist of 2-to-2 scattering, then $C[f]$ for some new species $X$ is defined as the sum over all such possible interactions involving $X$. If each interaction process is time-reversal invariant,

\begin{equation}
\label{eqn:collisionalint}
C[f_X] = \frac{1}{2} \displaystyle\sum_{X,i \rightarrow j,k} \int \left( \displaystyle\prod_{s=i,j,k} g_s \frac{d^3 p_s}{(2\pi)^3 2E_s} \right) (2\pi)^4 \delta^4(p) S \left| \mathcal{M} \right|^2 \Omega(f_X, f_i, f_j, f_k) \textrm{,}
\end{equation}

{\noindent}with the squared amplitude $|\mathcal{M}|^2$ averaged over the spins of both incoming and outgoing particles. The term $S$ corresponds to a symmetry factor whose value is $\frac{1}{2}$ when $j$ and $k$ are identical particles, to avoid overcounting of states in the phase space integral, and is $1$ otherwise. The $\Omega(\{f\})$ function is the phase space weighting term

\begin{equation}
\Omega(f_X,f_i,f_j,f_k) = f_j f_k (1 \pm f_X) (1 \pm f_i) - f_X f_i (1 \pm f_j) (1 \pm f_k) \textrm{,}
\label{eq:omega}
\end{equation}

{\noindent}where the $\pm$ term is $+$ for bosons (Bose enhancement) and $-$ for fermions (Pauli exclusion). The collision terms therefore couple together the Boltzmann equations for various particle species.

A detailed treatment of the full evolution of species in the early universe can be found in \cite{Bernstein}. For our purposes, the most important fact is that during non-equilibrium events, specifically the decoupling or annihilation of a species, the momentum dependence of the collision functional $C[f]$ can alter the phase space density of a decoupling species away from the standard thermal distributions. For these cases, a general phase space density must be numerically evolved in time to find the precise contribution to $g_*$ at lower temperatures. The focus of this work includes both decoupling and annihilation, necessitating our numerical treatment. 

So far we have treated the expansion of the universe as an independent process, but it is in fact coupled to the evolution of its particle content through the Einstein field equations. Assuming a flat, isotropic, and homogeneous universe, we obtain the Friedmann equations,

\begin{equation}
\begin{split}
H^2 &= \frac{8 \pi G}{3} \rho \textrm{,} \\
\frac{\partial \rho}{\partial t} &= -3H(\rho + P) \textrm{,}
\end{split}
\label{eq:friedmann}
\end{equation}

{\noindent}where $\rho$ and $P$ refer to the total energy density and pressure of the full particle content. The Boltzmann equations and Friedmann equations then combine to give a coupled set of integro-differential equations governing the full evolution of the early universe.


\subsection{Decoupling, recoupling, and the redistribution of entropy}

While a full solution to the Boltzmann and Friedmann equations is necessary to understand the detailed evolution of any species $X$ and its exact contribution to $g_*$, we can first gain a qualitative understanding by considering the approximation of instantaneous decoupling. Once we have developed this conceptual framework, we will then turn to more precise statements about the complete evolution of distribution functions.

In the instantaneous decoupling approximation, the point of decoupling can be found by comparing the rate of expansion $H$ to the rate of interaction $\Gamma_X$, defined as

\begin{equation}
\Gamma_X = \displaystyle\sum_{j,k \rightarrow X,i} \frac{n_j n_k}{n_X} \langle \sigma v \rangle_{j,k \rightarrow X,i} \textrm{,}
\end{equation}

{\noindent}where $\langle \sigma v \rangle$ is the thermally-averaged cross-section for any interaction $j,k \rightarrow X,i$. This average cross-section can be formally defined as

\begin{equation}
\langle \sigma v \rangle_{j,k \rightarrow X,i} = \int \left( \displaystyle\prod_{s=X,i,j,k} g_s \frac{d^3 p_s}{(2\pi)^3 2E_s} \right) (2\pi)^4 \delta^4(p) S \left| \mathcal{M} \right|^2 \frac{f_j f_k}{n_j n_k} (1 \pm f_X) (1 \pm f_i) \textrm{.}
\end{equation}

Note that the symmetry factor $S$ now includes an additional factor of $\frac{1}{2}$ if the initial state consists of identical particles, as well as the original $\frac{1}{2}$ for an identical-particle final state. The full set of thermally-averaged cross-sections can be related to the collisional term $C[f]$ via

\begin{equation}
\int g_X \frac{d^3 p_X}{(2\pi)^3 E_X} (2\pi)^4 \delta^4(p) C[f_X] S = \displaystyle\sum_{X,i,j,k} \left( n_j n_k \langle \sigma v \rangle_{j,k \rightarrow X,i} - n_X n_i \langle \sigma v \rangle_{X,i \rightarrow j,k} \right) \textrm{.}
\end{equation}

Conceptually, $\Gamma_X$ corresponds to the rate of production per particle for species $X$. As the universe expands, both $H$ and $\Gamma_X$ will decrease, though generically at different rates. If $\Gamma_X$ decreases more quickly than $H$, then it is possible for a species originally in equilibrium to `freeze out' and decouple from the remainder of the SM. 

Conversely, if $H$ decreases more quickly than $\Gamma_X$, a species originally out of equilibrium may actually recouple to the SM. In this case, however, $X$ will generically not have the same temperature as the SM, if it even has a well-defined temperature, prior to recoupling. Instead, the initial distribution will depend on any other particle content that could potentially couple to $X$, making this scenario very model-dependent.

In this framework, the point of instantaneous decoupling/recoupling is defined simply as the temperature at which $\Gamma_X = H$. It is common to assume that species are in full equilbrium prior to decoupling, then evolve freely immediately after freezing out. This approximate description is correct only when all relevant species are relativistic and originally in full equilibrium. However, if $X$ decouples during other nonequilibrium processes, such as nonrelativistic annihilation, the full set of Boltzmann equations must be used.

Once $T$ drops below the mass of any particle, that species begins to annihilate away, with the number density quickly falling to a negligible amount. The entropy of the annihilating species is redistributed amongst the remaining interacting species, such that the temperature of all remaining species decreases less quickly than would be the case in free expansion. If $X$ has decoupled from the SM prior to this annihilation, it will not participate in the resulting entropy redistribution, and therefore reaches a temperature lower than that of the SM following the annihilation.

To determine the impact of these entropy redistributions, we need to track the relativistic entropy density $s$ as a function of temperature. If the entropy density of all SM species in equilibrium (excluding $X$) was initially $s_0$ when $X$ instantaneously decoupled from the SM at temperature $T_0$, conservation of total entropy gives us the resulting temperature ratio following an entropy redistribution. This ratio can be expressed as a function of the entropy density $s$ of all species in equilibrium at any future temperature $T$,

\begin{equation}
\frac{T_X}{T} = \left( \frac{s/T^3}{s_0/T_0^3} \right)^{1/3} \textrm{.}
\end{equation}

In practice, because the entropy of annihilating species is only being distributed amongst relativistic species in full thermal equilibrium, it is much simpler and equivalent to instead use the relativistic degrees of freedom, rather than $s/T^3$, to calculate the ratio

\begin{equation}
\frac{T_X}{T} = \left( \frac{g_*^\mathrm{after}}{g_*^\mathrm{before}} \right)^{1/3} \textrm{,}
\end{equation}

{\noindent}where $g_*^\mathrm{before}$ and $g_*^\mathrm{after}$ are the relativistic degrees of freedom of all SM species in equilibrium immediately before and after the entropy redistribution. This decrease in relative temperature also decreases the $\Delta g_*$ due to $X$,

\begin{equation}
\Delta g_* = \Delta g_{*0} \left( \frac{g_*^\mathrm{after}}{g_*^\mathrm{before}} \right)^{4/3} \textrm{,}
\label{eq:deltagstar}
\end{equation}

{\noindent}where $\Delta g_{*0}$ is simply the initial contribution of $X$ to $g_*$ at $T_0$. For multiple entropy redistributions, the overall ratio $\frac{T_X}{T}$ can be found simply by multiplying together the ratios from each individual redistribution, giving the full contribution of $X$ to $g_*$.

Again, this discussion has made the simplifying assumption of instantaneous decoupling. In general, we cannot simply use comparisons of $\Gamma_X$ to $H$ to determine the exact evolution of the phase space density $f_X(t,E)$ if the species $X$ decouples during nonequilibrium processes. Our treatment must instead be made more precise by numerically solving the Boltzmann equation for $X$, as well as the Friedmann equations, which govern the evolution of the SM temperature $T(t)$ and the expansion scale factor $a(t)$. More details on our numerical treatment can be found in appendix \ref{sec:code}.

The evolution of a given model of new light species is determined by calculating the collision functional $C[f_X]$ in terms of the model parameters, such as the suppression scale $\Lambda$ of nonrenormalizable operators in an effective theory. This interaction term then governs the process of decoupling $X$ from the SM. Any SM annihilation and entropy redistribution that occurs after this decoupling reduces the change in effective degrees of freedom $\Delta g_*$ at the point of recombination. The contribution to $g_*$ for a specific model can be found by using the resulting $f_X$ near the point of recombination to calculate the energy density $\rho_X$. Solving this contribution in terms of generic couplings establishes a direct relationship between model parameters and $\Delta g_*$.

It is important to note that in this work we consider the effective field theory of each model at very low energies (as low as $\sim$ MeV). In order to match to any full UV theory which generates the operators in this effective theory, one should in principle treat operator couplings as Wilson coefficients and run these couplings from the high energy theory down to the scale of interest using the renormalization group. We assume that this running has already been done when we write down our effective operators, such that we are working with the matched coefficient.


\subsection{Relevant decoupling temperatures}
\label{sec:relevanttemps}

\begin{figure}[t]
\centering
\includegraphics[width=15cm]{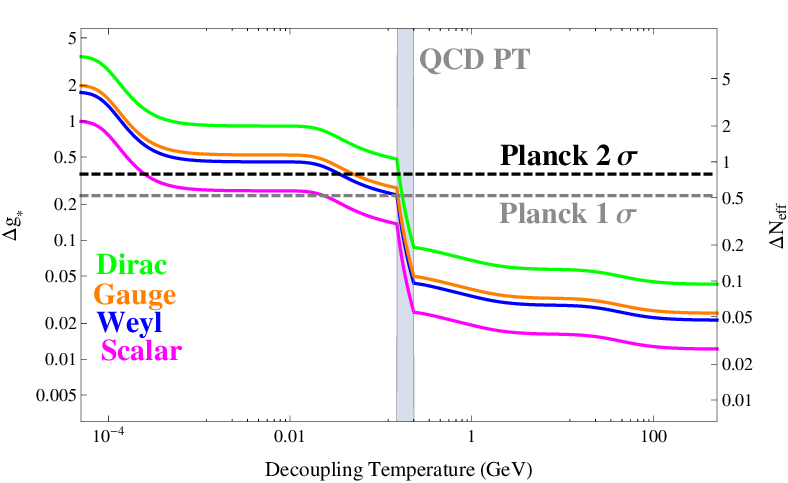}
\caption{Additional light degrees of freedom $\Delta g_*$ at recombination for a new light species as a function of the decoupling temperature (in the instantaneous decoupling approximation), calculated using eq.\ (\ref{eq:deltagstar}). The contribution of various particle species is shown, specifically a real scalar boson (magenta), a Weyl fermion (blue), a real gauge boson (orange), and a Dirac fermion pair (green). The dashed line indicates the current sensitivity of the Planck observational experiment \cite{Planck}. The gray region corresponds to the QCD phase transition, where the precise evolution of $g_*(T)$ for the SM is not well-understood. The provided values of $\Delta g_*$ should therefore only be interpreted qualitatively in that region.}
\label{fig:dgeffvsT}
\end{figure}

\begin{figure}[t]
\centering
\includegraphics[width=15cm]{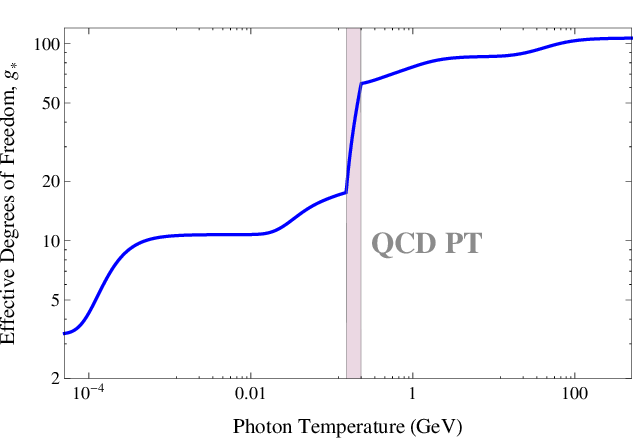}
\caption{Effective degees of freedom $g_*$ in the SM as a function of temperature. The gray region corresponds to the QCD phase transition, where the precise evolution of $g_*(T)$ is not well-understood. The provided values of $g_*$ should therefore only be interpreted qualitatively in that region.}
\label{fig:geffvsT}
\end{figure}

For new light species to currently be detectable with the CMB, they must decouple at low enough temperatures such that their contribution $\Delta g_*$ is within the experimental sensitivity of Planck \cite{Planck}. The full dependence of $\Delta g_*$ on the decoupling temperature for various particle types is shown in figure \ref{fig:dgeffvsT}. This functional dependence is calculated in the instantaneous decoupling approximation by using eq.\ (\ref{eq:deltagstar}) in combination with $g_*$ of the SM as a function of temperature, which is shown in figure \ref{fig:geffvsT}.

As we see in figure \ref{fig:dgeffvsT}, for a species to be within the sensitivity of Planck, it must decouple at temperatures $T \lesssim 200$ MeV, which corresponds to the approximate scale of the QCD phase transition (see \cite{Levkova:2012jd} and references therein for details). Prior to this point, quarks and gluons are the relevant degrees of freedom for the QCD sector, such that the total number of SM degrees of freedom is $g_* = 61.75$. As the universe cools to lower temperatures, the SM transitions to a regime where mesons and baryons are the appropriate degrees of freedom. Specifically, the relevant hadrons present below the QCD phase transition are pions and charged kaons, such that $g_* = 19.25$. This significant reduction in the degrees of freedom results from the rapid annihilation or decay of any more massive hadrons which may have formed during the transition. The QCD phase transition therefore corresponds to a large redistribution of entropy into the remaining degrees of freedom, such that any species which decouples from the SM prior to the transition will not contribute significantly to the CMB.

In principle, it is possible to discover species which decouple during the QCD phase transition, as those species could contribute values of $\Delta g_*$ above the experimental sensitivity. However, the precise details of this phase transition are not well-understood because of, e.g., strong coupling effects, and this transition is an area of active study (see \cite{Rischke,CasalderreyEtAl,LeupoldEtAl} and references therein). Consequently, we do not know how to make precise predictions for $\Delta g_*$ for species decoupling in this era. These computations are beyond the scope of our work, so we choose to restrict our focus to species which decouple after the QCD phase transition.

For new species which do decouple immediately after this point, the calculation of $\Delta g_*$ is sensitive to whether the species couples to leptons or to quarks. Species which couple solely to leptons have a straightforward decoupling process, as all relevant interactions are sufficiently weakly renormalized. Species which couple to quarks will then couple to pions and kaons, whose couplings can be strongly renormalized. We must restrict ourselves to quark and meson couplings which involve conserved currents, as these are then protected against strong renormalization effects. For this set of couplings, we can still make precise predictions for the contribution of new light species which couple to quarks, even at temperatures immediately below the QCD phase transition.


\subsection{Big Bang Nucleosynthesis}

Most models which include additional light degrees of freedom will have other model-dependent constraints, such as those from collider signals or various astrophysical observations. Arguably the most important model-independent bound other than that of the CMB is that placed by Big Bang Nucleosynthesis (BBN). The measurement of the primordial relic abundance of light elements formed by BBN provides an independent probe of new light species, although at times earlier than recombination. While here we only give a brief summary of the relevant aspects of BBN, an excellent introduction to the topic can be found in \cite{Steigman07,PospelovEtAl,IoccoEtAl}.

The resulting abundances of the light elements, particularly helium-4 ($^4$He), are sensitive to the number density of neutrons at the start of BBN. When neutrons and protons were in full equilibrium, the number of neutrons relative to that of protons continued to fall due to their mass splitting. The neutron abundance is then determined by the point at which the weak interactions, which interconvert protons and neutrons, freeze out. A larger expansion rate results in earlier freezeout, which in turn leads to a larger number of neutrons and therefore more $^4$He.

The precise value of $H$ at the time of BBN, which would be increased by the presence of additional light species, therefore determines the relic abundance of $^4$He. This abundance is often expressed in terms of the so-called `helium mass fraction'

\begin{equation}
Y_P \equiv \frac{4 n_{He}}{n_H + 4 n_{He}} \textrm{.}
\end{equation}

Observational determinations of $Y_P$ therefore provide another means of constraining the relativistic energy density of the early universe, though it is important to remember that these constraints apply at a different period of time than those placed by direct CMB measurements of $g_*$. The SM prediction for the primordial helium abundance is $Y_P = 0.2487 \pm 0.0006$ \cite{Steigman07}, and this prediction can be tested by both extracting the primordial abundance from direct observations of the modern helium abundance and observing the effects of $Y_P$ on CMB anisotropies.

Multiple primordial helium extractions have yielded results near $Y_P = 0.240 \pm 0.006$ \cite{Steigman07}, which are consistent with SM predictions, but two recent observational studies have indicated a higher abundance of $Y_P = 0.2565 \pm 0.0010$ (stat) $\pm 0.0050$ (syst) \cite{IzotovEtAl} and $Y_P = 0.2561 \pm 0.0108$ \cite{AverEtAl}, which are consistent with a larger rate of expansion. This in turn allows for the presence of new light species. In addition, combined CMB constraints from SPT and WMAP are consistent with $Y_P = 0.296 \pm 0.030$ \cite{KeislerEtAl}, and combined results from Planck and WMAP are consistent with $Y_P = 0.266 \pm 0.021$ \cite{Planck}. These results are therefore currently incapable of either completely confirming or excluding the existence of new light species, but instead increase the importance of the precision CMB measurements of $g_*$ possible with future experiments.

Lastly, it is important to note that there is tension between the SM prediction and observational determinations of the abundance of lithium-7 ($^7$Li), with a lower observationally inferred primordial $^7$Li abundance than that predicted by BBN. Unfortunately, this discrepancy is not immediately remedied simply by the presence of new light species, and the detailed model-building necessary to address this tension is beyond the scope of this paper. However, the $^7$Li problem does present another exciting opportunity for the possible discovery of new physics \cite{Pospelov:2010cw,Fields,KusakabeEtAl,KohriEtAl,CyburtEtAl}.


\subsection{Summary}
\label{sec:summary}

We have now introduced the framework necessary for the remainder of this paper. The focus of this work is the effects of light species in BSM theories on the CMB, which we determine by computing the energy density of the new light species at recombination. We specifically concern ourselves with species which were in thermal equilibrium with the SM and then decouple after the QCD phase transition, potentially during the annihilation of a SM species. Any species which decouples from the SM before the QCD phase transition cannot be probed by the Planck satellite, as its energy density is much smaller than that of the SM. The energy density of light species is calculated by numerically solving the coupled Boltzmann and Friedmann equations, found in eqs.\ (\ref{eq:boltz}) and (\ref{eq:friedmann}), in order to compute the potentially nonthermal distribution function of the new species. The distribution function immediately following decoupling can then be used to calculate the energy density at recombination, which determines $g_*$ using eq.\ (\ref{eq:rhorel}).


\section{Models}
\label{sec:models}

In this section, we consider the set of models which can contribute to the CMB measurement of $g_*$\footnote{We only consider models with light degrees of freedom. It is possible to construct models where heavier species mimic the effects of light degrees of freedom through a nonzero presure resulting from non-equilibrium distribution functions \cite{Hooper:2011aj}.}, mainly restricting ourselves to models where the additional degrees of freedom were in thermal equilibrium immediately following the QCD phase transition\footnote{There are models where out-of-equilibrium effects such as decays generate a contribution to $g_*$ \cite{GonzalezGarciaEtAl,HasenkampEtAl,BlennowEtAl,MenestrinaEtAl, FischlerEtAl,Bjaelde:2012wi,Cicoli:2012aq,GrafEtAl12,GrafEtAl13,HigakiEtAl}, but no generic model-independent statements can be made about such scenarios, so we do not consider them in this work.}. Such models must either contain new species with mass $\lesssim$ eV or alter the neutrino energy density. While there are a very large number of possible models one could write down, we choose to restrict ourselves to those which are both minimal and natural.

We consider a model to be minimal if it contains the smallest possible hidden sector in the low-energy theory. In particular, this restricts our discussion to models of elementary particles, ignoring the possibility of light composite states. We then direct our attention to the low-energy effective field theory (EFT) and ignore any additional particle content which may arise at higher energies, as these are irrelevant for our calculations.

For this work, we define naturalness as technical naturalness. We therefore require that the size of quantum corrections not exceed the size of the physical observables in the theory, i.e. $|\frac{\delta \lambda}{\lambda}| < 1$ for all parameters $\lambda$, as large corrections require an artificial fine-tuning of parameters.

A large number of potential models of light species are unnatural, due to large corrections to the mass of that new species. There are two predominant methods of suppressing quantum corrections to a particle's mass. The first method is the introduction of an additional symmetry which prohibits the existence of a mass term for that species. The second option is the use of strong dynamics in a hidden sector to generate large anomalous dimensions for mass terms, such that those terms become irrelevant operators, giving rise to a vanishing mass in the low-energy EFT. However, most models of the latter type tend to contain a relatively rich spectrum, violating our minimality principle. Although this is an interesting direction for future research, it is outside the class of models we consider. We therefore focus solely on theories of light species which contain a protective symmetry.

The classes of possible new light species can be divided up by spin, as this restricts the protective symmetries available. We progress through each possible case, from spin-0 to spin-2, considering all minimal, natural models. For each model, we then scan over all allowed couplings, numerically solving the Boltzmann and Friedmann equations to calculate the full process of decoupling for any species which decouples during SM entropy redistributions. The details of our numerical approach can be found in appendix \ref{sec:code}. Using the resulting distribution function after the decoupling of our new light species, we then calculate the energy density at recombination, which is reported as the contribution to $g_*$ as a function of the coupling parameters of the theory. This calculation of $\Delta g_*$ is specifically done in the massless limit, and is accurate for new species with $m \ll$ eV. In subsection \ref{sec:massive}, we briefly discuss the potential effects of non-negligible masses.

As discussed in subsection \ref{sec:relevanttemps}, each new light species must also decouple after the QCD phase transition in order to be constrained by Planck, which limits the dimensionality of the operators we choose to consider. If our new species couples to the SM with an operator of scaling dimension $d$, the operator is suppressed by $\Lambda^{4-d}$, where $\Lambda$ is the approximate cutoff scale of the EFT. Dimensional analysis then indicates that, given independent experimental constraints, only operators of dimension $d \lesssim 6$ will be able to maintain equilibrium between a new species and the SM until after the QCD phase transition.

Finally, we discuss possible extensions to the SM which do not contain new light species, but instead alter the neutrino distribution, through such means as decay or neutrino asymmetry. These models then enhance the neutrino energy density relative to SM predictions, leading to an increase in $g_*$.


\subsection{Spin-$0$: Goldstone boson}

The first possibility for new light species is a spinless scalar boson. However, the mass of any new scalar particle is generically sensitive to quantum contributions resulting from interactions. While supersymmetry could potentially preserve the naturalness of scalar masses, the observed particle spectrum indicates that any couplings between the SM and new light scalars would mediate supersymmetry-breaking mass terms significant enough to require fine-tuning. The only viable symmetry which can protect the mass term of such light scalar bosons is then a shift symmetry, $\phi \rightarrow \phi + \epsilon$. This is precisely the symmetry present in the Goldstone modes of a spontaneously broken global symmetry. In the limit of an exact global symmetry, the mass of the corresponding Goldstone boson is restricted to be zero, with any quantum corrections forbidden by the symmetry. Even if the symmetry is inexact, the mass of the pseudo-Goldstone is proportional to the symmetry-breaking terms in the original Lagrangian, rather than the cutoff of the effective theory. We therefore restrict ourselves to the study of Goldstone bosons, as other theories of light scalars are generically tuned and unnatural. These particles arise in many theories, such as the QCD axion and the so-called `String Axiverse' of string compactifications \cite{Arvanitaki:2009fg}.

Since we only discuss thermodynamics after the QCD phase transition, the only allowed interactions are those with leptons, mesons, baryons, and the photon. In this low energy effective theory, any combination of such couplings may conceivably be allowed. We explore all of these possibilities, finding that current collider and astrophysical bounds are such that almost all scenarios are excluded, and minimal models of Goldstone bosons must have a negligible impact on $g_*$. There are small corners in (flavor-dependent) parameter space which are still viable and in which they could in principle have a small but non-negligible impact on the effective number of relativistic degrees of freedom.  We conclude that unless we are very lucky, the addition of a natural massless or near massless scalar will have, at best, a tiny impact on the CMB and thus would require significant advances in our ability to measure $g_*$.

First, we consider couplings to leptons. Due to the shift symmetry, any coupling between an exact Goldstone boson and SM fermions must only contain derivatives of the field $\phi$. We parameterize our effective field theory as

\begin{equation}
\mathcal{L} \supset \, \frac{1}{2} (\partial_\mu \phi)^2 + \frac{\partial_\mu \phi}{2 \Lambda} \psi_L^\dagger \bar{\sigma}^\mu \psi_L 
+ \frac{\partial_\mu \phi}{2 \Lambda} \psi_R^{c\dagger} \bar{\sigma}^\mu \psi_R^{c} + h.c.
\end{equation}

Using identities found in \cite{Dreiner:2008tw}, we can also write our Lagrangian in Dirac notation, resulting in

\begin{equation}
\mathcal{L} \supset \, \frac{1}{2} (\partial_\mu \phi)^2 - \frac{\partial_\mu \phi}{\Lambda} \bar{\Psi} \gamma^\mu \gamma^5 \Psi + h.c.
\end{equation}

{\noindent}In this form, it is simple to see that the interaction is specifically a derivative coupling between $\phi$ and the axial current of $\Psi$. One might suspect that some theories could potentially generate a similar coupling between $\phi$ and the vector current for $\Psi$. However, any interaction of that form must vanish due to vector current conservation. The conservation of the axial current is broken by the mass term for $\Psi$, meaning that the axial coupling does not similarly vanish. However, this does imply that any interaction rate involving the axial coupling is necessarily proportional to the fermion mass $m$, and thus vanishes in the $m \rightarrow 0$ limit.

In simple UV completions of this effective field theory, the couplings of $\phi$ to the SM are flavor-blind. More sophisticated UV model-building could potentially result in flavor-specific couplings. However, a flavor-specific basis generically leads to interactions which mix generations. There are greatly restrictive constraints coming from flavor physics, as we shall discuss briefly below.

\begin{figure}[t]
\centering
\includegraphics[width=0.4\textwidth]{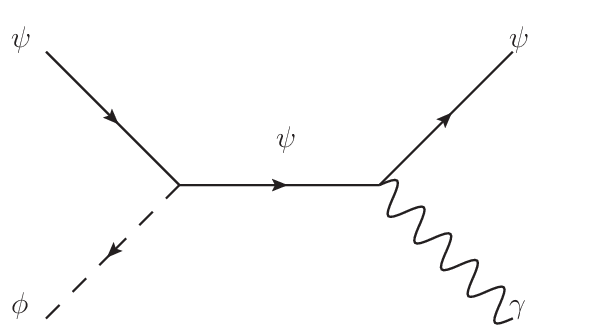}
\caption{Dominant interaction process for the Goldstone-lepton coupling.}
\label{fig:op02}
\end{figure}

Due to the $\Lambda$ suppression of the derivative couplings, the interaction rate between $\phi$ and leptons will be dominated by processes which only involve one Goldstone interaction term, shown in figure \ref{fig:op02}. Note that, as this dominant process involves the emission/absorption of a photon, the interaction rate $\Gamma_\phi$ has no dependence on the coupling between $\phi$ and neutrinos. Because of this, the only relevant lepton interactions for $\phi$ are those with electrons and muons.

In the relativistic limit, dimensional analysis would expect the interaction rate between $\phi$ and a SM lepton $\Psi$ to scale as $\Gamma_\phi \sim \frac{T^3}{\Lambda^2}$. However, the broken axial symmetry for $\Psi$ restricts the interaction rate to take the form $\Gamma_\phi \sim \frac{m^2 T}{\Lambda^2}$. The expansion rate will therefore drop more quickly than the interaction rate as the universe expands. This implies that $\phi$ could have been out of thermal equilibrium after the time of global symmetry breaking, and come back {\it into} thermal equilibrium with the SM leptons at some point before the leptons annihilated.

As the universe cools to temperatures comparable to the relevant lepton mass, this simplified form for the interaction rate will be substantially modified and needs to be computed numerically. The interaction rate $\Gamma_\phi$ will begin to drop rapidly as the leptons annihilate away, redistributing their entropy amongst the remaining coupled species. If $\Lambda$ is very large and substantially suppresses $\Gamma_\phi$, $\phi$ will not have recoupled by the time the leptons annihilate, meaning that $\phi$ will forever remain out of equilibrium. Additional couplings beyond the lepton-only couplings we consider in this paragraph would be needed in order to have SM-$\phi$ interactions. For each lepton, there is then some maximum $\Lambda$ for recoupling. Any Goldstone boson with a larger $\Lambda$ will not be reheated by the entropy redistribution and therefore cannot substantially contribute to $g_*$ at recombination.

The distribution function for $\phi$ prior to recoupling is dependent on the original process of decoupling at high energies, which is then sensitive to details of the UV theory, including the relative timings of global symmetry breaking and inflation. It is then impossible to make fully model-independent predictions for the contribution of $\phi$ to $g_*$ in the case where $\phi$ only couples to leptons. However, in a large class of models, $\phi$ will also couple to quarks and photons, which results in qualitatively different evolution.

For the case where $\phi$ has similar couplings to quarks, we must examine the resulting interactions between $\phi$ and mesons, specifically pions and charged kaons, since we consider temperatures below the QCD phase transition. At these temperatures, however, the number density of kaons will be much lower than that of pions, such that any $\phi$-kaon interactions will be subdominant. We can then simply focus on those couplings which involve pions. Following \cite{GeorgiEtAl}, the original coupling of $\phi$ to quarks can be rewritten in terms of the axial quark current. After the phase transition to mesons, interactions with the quark current are replaced by those with the axial pion current, which is safe from QCD renormalization effects. The full Lagrangian can be expanded to leading order in ${\Lambda}$ and $f_\pi$ and subsequently studied, and depends on the details of the flavor structure in the UV. We assume flavor-blind couplings, as flavor-specific couplings in the UV do not alter our predictions for the thermodynamic properties of $\phi$, but such couplings must obey additional constraints coming from flavor physics. With this assumption, we find that those terms which dominate the interaction rate between $\phi$ and pions are

\begin{equation}
\label{eqn:piongoldstone}
\mathcal{L} \supset \, - \frac{2r_m}{3f_\pi \Lambda} \pi^+ \pi^- \partial_\mu \phi \partial^\mu \pi^0 + \frac{r_m}{3f_\pi \Lambda} \pi^0 \pi^+ \partial_\mu \phi \partial^\mu \pi^- + \frac{r_m}{3f_\pi \Lambda} \pi^0 \pi^- \partial_\mu \phi \partial^\mu \pi^+ \textrm{.}
\end{equation}

We have defined the ratio $r_m=\frac{m_d-m_u}{m_d+m_u}$, where $m_u$ and $m_d$ are the up and down current-quark masses. We use the approximate value $r_m = 1/3$, based on lattice QCD calculations \cite{BeringerEtAl}, as well as the convention $f_\pi = 93$ MeV. Interactions of this type will potentially keep the Goldstone boson $\phi$ in thermal equilibrium until the pions fully annihilate and redistribute their equilibrium, depending on the suppression scale $\Lambda$. The corresponding interaction rates decrease more rapidly than the expansion rate, leading to the freezing out of the $\phi$-$\pi$ interactions.

Finally, there can be couplings of $\phi$ to photons via operators of the form 

\begin{equation}
\mathcal{L} \supset - \frac{e^2}{32 \pi^2 \Lambda_{\gamma}} \phi F^{\mu \nu} \tilde{F}_{\mu \nu} \textrm{.}
\end{equation}

This operator arises because the axial symmetry in question can be anomalous. This $\Lambda_\gamma$ is not necessarily precisely the same as the $\Lambda$ which couples $\phi$ to SM fermions, though their orders of magnitude are similar in a large number of UV completions. This is because the operator can be induced by loops of SM fermions. The additional loop factor in the parameterization of $\Lambda_\gamma$ is present because in these cases, the operator appears in the Lagrangian suppressed by a loop factor relative to the fermion couplings. As mentioned earlier, depending on the UV structure of the model, this operator may or may not be present in the low-energy theory. Similar to pion couplings, this operator gives rise to a rate such that $\phi$-$\gamma$ interactions freeze out as we go to lower temperatures.

We now outline the constraints on these scenarios, working in a general framework with no assumptions regarding the operator or flavor structure of couplings in the UV. The bounds are best stated in terms of the effective operators

\begin{equation}
\mathcal{L} \supset - \frac{\partial_\mu \phi}{\Lambda_f} \bar{\Psi}_f \gamma^\mu \gamma^5 \Psi_f - \frac{e^2}{32 \pi^2 \Lambda_\gamma} \phi F^{\mu \nu} \tilde{F}_{\mu \nu}
\textrm{,}
\end{equation}

{\noindent}where $\Psi_f$ can either be a charged lepton or the proton. The strongest bounds for these models come primarily from observations of stellar and supernova cooling, which will also greatly constrain other models within this work. The production of new light species which interact weakly enough to escape the interior of a star provides an efficient energy loss mechanism, affecting both stellar cooling and evolution. Comparison of SM predictions to astrophysical observations then provides a strong constraint on the interactions of such new species. The resulting constraints for Goldstone interactions are

\begin{equation}
\begin{split}
\Lambda_e &\gtrsim 2.9 \times 10^9 \textrm{ GeV,} \\
\Lambda_p &\gtrsim 3.5 \times 10^9 \textrm{ GeV,} \\
\Lambda_\gamma &\gtrsim 1.2 \times 10^7 \textrm{ GeV.}
\end{split}
\end{equation}

More details about these bounds can be found in \cite{Raffelt96,Raffelt99,Raffelt06,Cicoli:2012sz}. The relationship between the effective proton scale $\Lambda_p$ and the UV quark coupling scale $\Lambda \equiv \Lambda_q$ present in eq.\ (\ref{eqn:piongoldstone}) depends on phenomenological parameters in the baryon chiral Lagrangian, as well as details of the UV theory. However, $\Lambda_p$ and $\Lambda_q$ are related by an $\mathcal{O}(1)$ number. Consequently, we use the conservative bound $\Lambda_q \gtrsim 5 \times 10^8$ GeV.

In addition, there are constraints on the set of off-diagonal operators schematically of the form $\frac{1}{\Lambda_{\mu e}} \partial_\mu \phi \bar{\mu} \gamma^\mu \gamma^5 e$ coming from $\mu \rightarrow e+ \slashed{E}$ \cite{BeringerEtAl}. These bounds restrict

\begin{equation}
\Lambda_{\mu e} \gtrsim 1.6 \times 10^9 \textrm{ GeV.}
\end{equation}

These off-diagonal operators' contributions to early universe thermodynamics are not significantly different from that of muon couplings during the era following the QCD phase transition. Consequently, we do not consider this case to be qualitatively distinct from the case with muon couplings, but considerably more constrained, and so we do not consider these operators further.

Finally, there are direct constraints on $\Lambda_\mu$ also coming from observations of supernovae. We take the average temperature within the core of a supernova to be $T \approx 30$ MeV \cite{Raffelt99}, which allows for the presence of a non-negligible muon abundance. We can therefore apply the same cooling bounds to muon couplings, with a small suppression due to the lower muon number density. Based on \cite{Raffelt99}, we calculate the approximate bound

\begin{equation}
\Lambda_\mu \gtrsim 2.0 \times 10^6 \textrm{ GeV.}
\end{equation}

\begin{figure}[t]
\centering
\includegraphics[width=15cm]{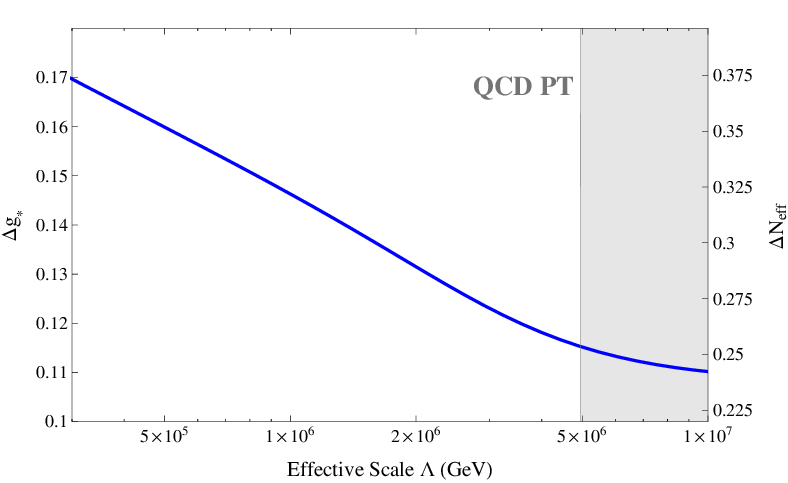}
\caption{$\Delta g_*$ due to a single Goldstone boson which interacts with only pions. The contribution to $g_*$ at recombination is given as a function of the effective scale $\Lambda$, which suppresses this interaction. The gray region for $\Lambda \gtrsim 5 \times 10^6$ GeV corresponds to models which decouple during the QCD phase transition. The provided values of $\Delta g_*$ should therefore only be interpreted qualitatively in that region. Supernova and star cooling constraints on this scenario limit $\Lambda \gtrsim 10^9$ GeV, and so this plot demonstrates that the Goldstone must have decoupled during or before the QCD phase transition.}
\label{fig:goldstone-pionly-results}
\end{figure} 

\begin{figure}[t]
\centering
\includegraphics[width=15cm]{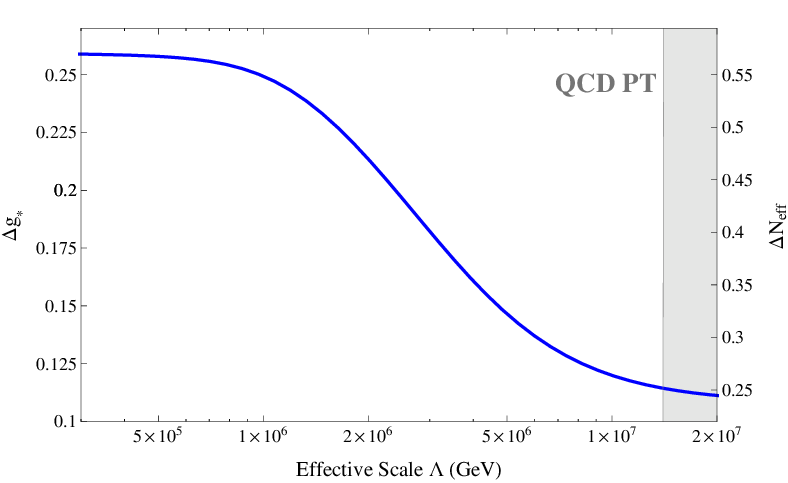}
\caption{$\Delta g_*$ due to a single Goldstone boson which interacts with at least pions and muons. The contribution to $g_*$ at recombination is given as a function of the effective scale $\Lambda$, which suppresses this interaction. The blue-gray region for $\Lambda \gtrsim 1.5 \times 10^7$ GeV corresponds to models which decouple during the QCD phase transition. The provided values of $\Delta g_*$ should therefore only be interpreted qualitatively in that region. Supernova and star cooling constraints on this scenario limit $\Lambda \gtrsim 10^9$ GeV, and so this plot demonstrates that the Goldstone must have decoupled during or before the QCD phase transition.}
\label{fig:goldstone-pimu-results}
\end{figure}

In order to consider the general list of all possible models, we present our results for each interaction separately. For a large number of models, multiple such interactions will be present, such that these results will be even more restrictive.

{\bf Electrons/Photons:} Electron interaction rates are suppressed by $\frac{m_e}{T}$, and photon interaction rates are suppressed by the loop factor $\frac{e^2}{32\pi^2}$, such that these two heavily-constrained interactions do not play a role in the thermal evolution of Goldstone bosons.

{\bf Pions:} For Goldstone bosons to be in thermal equilibrium with pions and receive any of the pion entropy redistribution, the coupling suppression scale must be $\Lambda \lesssim 5\times 10^6$ GeV. This is illustrated in figure \ref{fig:goldstone-pionly-results}. The maximum possible $\Lambda$ necessary is far below the bound on $\Lambda$ quoted above, and therefore the decoupling of the Goldstone must have happened during or before the QCD phase transition, making the Goldstone not a viable candidate for a contribution to $g_*$ in theories containing only pion interactions.

{\bf Muons:} In the case of muon-only couplings, it is not possible to give well-defined initial conditions for the Goldstone boson distribution function just prior to the recoupling of the Goldstone boson to muons. For all reasonable initial configurations of the Goldstone distribution function, the maximum contribution possible would result from thermalization of the Goldstone bosons with muons, leading to a contribution of $\Delta g_* = 0.26$, or $\Delta N_{eff} = 0.57$. We plan to pursue more precise predictions in future work.

However, if these couplings are present in conjunction with couplings to pions, then it is possible to study the decoupling of Goldstones from the SM, as the Goldstones had been in thermal equilibrium in the era leading up to muon annihilation. In order for a Goldstone to have received any entropy at all from SM annihilations following the QCD phase transition, it must have coupled with $\Lambda < 1.5 \times 10^7$ GeV. This is illustrated in figure \ref{fig:goldstone-pimu-results}. While such couplings are allowed for muon interactions, this range is below the pion bounds quoted above, and therefore the Goldstone is not a viable candidate for a contribution to $g_*$ in this scenario.

To summarize, there are no parts of the minimal, natural parameter space where the Goldstones had been in thermal equilibrium with the SM through the QCD phase transition which do not directly conflict with bounds coming from star and supernova cooling. As such, the effects of Goldstone bosons on the CMB in the predictive part of the parameter space are well below the sensitivity of the Planck satellite. One can, however, have couplings to only the muon with $\Lambda$ in the narrow window between $2.0 \times 10^6$ GeV and $1.5 \times 10^7$ GeV, and still obtain a nontrivial contribution to $\Delta g_*$, though it is not possible to give well-defined, model-independent initial conditions for the Goldstone distribution function in this scenario. Therefore, the only viable set of theories must contain a highly specific hierarchy of couplings, such that interactions with muons are much stronger than those with other SM fields present after the QCD phase transition, without the generation of significant off-diagonal couplings.


\subsection{Spin-$\frac{1}{2}$: Light fermion}

Natural models of light spin-$\frac{1}{2}$ fermions are made more easily than those containing light scalar bosons. This naturalness can arise due to chiral symmetry, which corresponds to a rotation of the field by an arbitrary phase, $\chi \rightarrow e^{i \alpha} \chi$. This symmetry permits any fermion gauge and kinetic terms, but forbids Majorana mass terms. Even if chiral symmetry is explicitly broken by the presence of a small fermion mass, corrections to this mass parameter are in general proportional to the original value, eliminating the need for any fine-tuning. Similarly, Dirac mass terms can be protected by an axial symmetry.

Because of this protective symmetry, there are many allowed interactions for light fermions. The possible models include interactions with SM gauge bosons, either through direct gauge couplings or dipole moments, as well as interactions with SM fermions through effectively pointlike operators, which result from the exchange of heavy intermediary particles.


\subsubsection{Gauge interactions}

One possibility is that a new light fermion $\chi$ is charged under the SM gauge groups. The coupling strength of a fermion in any representation of $SU(3)_C$ or $SU(2)_L$ is completely fixed by the representation theory of these groups. While $\chi$ could na\"{i}vely have any value of hypercharge, the prospect of gauge unification indicates that hypercharge values are also discrete. Any new fermion in non-trivial representations of the SM gauge groups will therefore couple with the same strength as the SM fermions. Light species which possess electromagnetic or color charge are completely excluded. The only remaining option is a neutral fermion, which must couple to the $Z$, but these light fermions are excluded by measurements of the $Z$-width \cite{BeringerEtAl}. As such, light fermions in any non-trivial representation of the SM gauge groups are excluded as potential candidates for contributions to $g_*$.

However, if $\chi$ instead coupled to some new gauge boson, kinetic mixing between this new field and the SM gauge bosons would lead to mixing-suppressed SM gauge couplings for $\chi$. Any such `millicharged' light fermion therefore requires the existence of a new gauge boson, which would also contribute to $g_*$. We consider the details of new gauge bosons and the resulting millicharged interactions in subsection \ref{sec:vector}.


\subsubsection{Dipole and anapole moments}

While a new fermion cannot carry SM charges, $\chi$ could still interact via dimension-5 dipole moment operators. The only nontrivial dipole interactions between $\chi$ and SM gauge bosons are those with the hypercharge gauge boson, which are of the form

\begin{equation}
\mathcal{L} \supset - \, \frac{1}{\Lambda} B_{\mu \nu} \chi_L \sigma^{\mu \nu} \chi_R^c + h.c. \textrm{,}
\label{eqn:chidipole}
\end{equation}

{\noindent}where the structure of these operators is such that we must introduce two new Weyl fermions, $\chi_L$ and $\chi_R^c$. These interactions can arise from loops involving heavy charged intermediaries, whose mass and couplings set the dipole moment scale $\Lambda$.

However, the charged intermediary loops that generate this operator necessarily preserve only the vector $U(1)$ global symmetry of $\chi$, which is precisely the symmetry structure allowed by a Dirac mass term $m \chi_L \chi_R^c$. Therefore, any UV completion reducing to the theory containing the Lagrangian terms of eq.\ (\ref{eqn:chidipole}) must also allow for a Dirac mass term. It is not apparent how to create a UV completion of this model which induces only a dipole term corresponding to large mass scales, while generating the Dirac mass $\lesssim$ eV in a natural fashion. Experimental constraints from star cooling observations \cite{Raffelt12} currently limit $\chi$ dipole moments to

\begin{equation}
\Lambda \gtrsim 10^9 \textrm{ GeV.}
\end{equation}

Due to the resulting large separation of scales in this highly constrained EFT, we do not consider a theory with new light species possessing a SM dipole moment to be a viable, natural candidate for a contribution to $\Delta g_*$.

A similar interaction term corresponds to the anapole moment and charge radius operators, which are of the form

\begin{equation}
\mathcal{L} \supset - \, \frac{1}{\Lambda^2} \chi^\dagger \bar{\sigma}^\mu \chi \partial^\nu B_{\mu \nu}  \textrm{.}
\end{equation}

Such interactions are dimension-6 and only require the existence of a single new Weyl fermion $\chi$. New species with such interactions were discussed in the context of dark matter in \cite{HoEtAl2}. Unlike dipole moments, such anapole moment interactions do not break chiral symmetry and are therefore compatible with new light or massless species, not just nonrelativistic dark matter.

Assuming vanishing boundary terms, the anapole interaction can be rewritten as

\begin{equation}
\mathcal{L} \supset \, \frac{1}{\Lambda^2} \partial_\mu \left( \chi^\dagger \bar{\sigma}^\mu \chi \right) \partial_\nu B^\nu + \frac{1}{\Lambda^2} \chi^\dagger \bar{\sigma}^\mu \chi \partial^2 B_\mu \textrm{,}
\end{equation}

{\noindent}which then results in couplings between $\chi$ and both the photon and $Z$. Similar to the case of Goldstone bosons, processes involving the first interaction term will be proportional to $m_\chi$, as this interaction involves the divergence of a current which is conserved in the limit $m_\chi \rightarrow 0$. As $m_\chi \ll T$ for all cases we consider, such processes are greatly suppressed and this particular interaction is irrelevant to our discussion.

The second interaction term is not similarly suppressed but instead has the form of a gauge coupling with additional momentum dependence. The dominant process involving this interaction is the exchange of a photon between $\chi$ and SM fermions. In such processes, the extra powers of momentum in this operator will cancel with those of the photon propagator, resulting in an amplitude of the same form as four-fermion interactions between $\chi$ and the SM. While the full models generating four-fermion interactions are very different from those which generate anapole moments, the phenomenology and the resulting bounds on the suppression scale $\Lambda$ will be very similar for both models. The results for four-fermion interactions, which are discussed in the following subsection, can therefore easily be applied to models involving anapole moment interactions.


\subsubsection{Four-fermion interactions}
\label{sec:4fermion}

Another possibility for EFT interactions of light fermions is the dimension-6 couplings of a single Weyl fermion $\chi$ or a Dirac pair of fermions $\mathrm{\bf X}$ to SM fermions. Such couplings can arise due to the exchange of a massive scalar or vector boson. Spontaneously broken gauge symmetries, which generate such massive interactions, are present in a large class of theories. Two well-motivated examples are the addition of light sterile neutrinos which couple to the SM via a new massive gauge boson $Z'$, corresponding to a spontaneously broken $U(1)$ \cite{AnchordoquiEtAl,SolagurenBeascoaEtAl,AnchordoquiGoldberg,EngelhardtEtAl}, and theories where the axino, the supersymmetric partner of the axion, remains light and interacts with the SM via other heavy superpartners.

As we will see below, light Weyl fermions with dimension-6 couplings are strong candidates for significant contributions to $g_*$. Interactions suppressed by scales $\Lambda \sim 2$ TeV will keep new species in equilibrium until after the QCD phase transition, leaving such species with a detectable energy density at recombination. The strongest independent bounds on such models are placed by collider experiments, which will continue to probe the relevant parameter space. These theories will then potentially be discovered or fully excluded with the LHC.

In Dirac notation, the possible four-fermion operators present after electroweak symmetry breaking (EWSB) take four forms,

\begin{equation}
\begin{split}
\frac{1}{\Lambda^2} \bar{\mathrm{\bf X}} \mathrm{\bf X} \bar{\Psi} \Psi & \hspace{2em} \textrm{(Scalar),} \\
\frac{1}{\Lambda^2} \bar{\mathrm{\bf X}} \gamma^5 \mathrm{\bf X} \bar{\Psi} \gamma^5 \Psi & \hspace{2em} \textrm{(Pseudoscalar),} \\
\frac{1}{\Lambda^2} \bar{\mathrm{\bf X}} \gamma^\mu \mathrm{\bf X} \bar{\Psi} \gamma_\mu \Psi & \hspace{2em} \textrm{(Vector),} \\
\frac{1}{\Lambda^2} \bar{\mathrm{\bf X}} \gamma^\mu \gamma^5 \mathrm{\bf X} \bar{\Psi} \gamma_\mu \gamma^5 \Psi & \hspace{2em} \textrm{(Axial),}
\end{split}
\end{equation}

{\noindent}where $\Psi$ corresponds to any SM fermion and the suppression scale $\Lambda$ arises from the mass and couplings of the exchanged intermediary. We instead discuss the couplings of a Weyl fermion $\chi$ below, as our results can simply be scaled by a factor of $2$ to account for the two fermions in the Dirac case. As all four operators are dimension-6, the interaction rate will drop more quickly than $H$, leading to the decoupling of $\chi$ from the SM as the universe cools.

Couplings of this new fermion to quarks can induce couplings to pions in the low-energy theory. However, any interaction arising from the scalar or pseudoscalar operators will not be protected against strong renormalization effects, such that we cannot make precise theoretical predictions for $\Delta g_*$. If such a species is independently discovered, potentially in collider experiments, and these interactions are precisely determined, a detailed calculation could then be performed. In addition, the vector or axial interactions are such that mesons will have no charge under such couplings, with no induced pion couplings in the EFT\footnote{Vector or axial interactions between pions and $\chi$ would result from models with couplings which are not flavor-blind. Such couplings can only arise from the spontaneous breaking of nonabelian gauge groups which do not commute with flavor symmetry. Such models require a significantly larger particle content, thus violating our minimality requirement.}.

\begin{figure}[t]
\centering
\includegraphics[width=15cm]{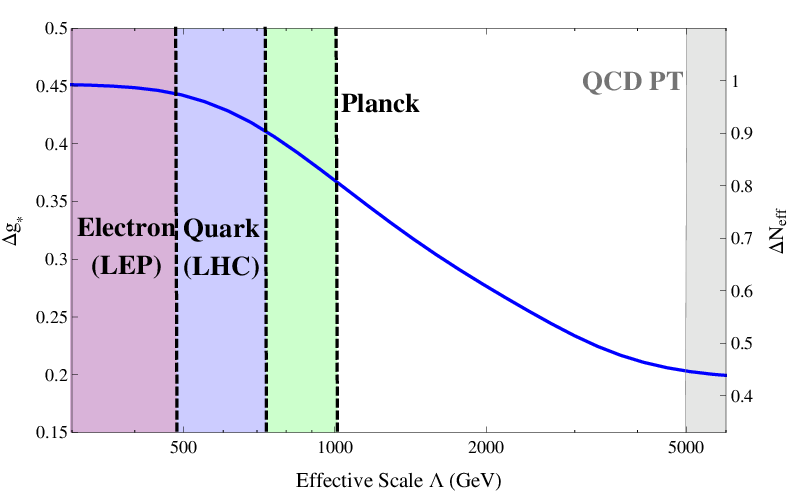}
\caption{$\Delta g_*$ due to a single Weyl fermion which interacts with the SM via the exchange of a massive vector boson. The contribution to $g_*$ at recombination is given as a function of the effective scale $\Lambda$, which suppresses this interaction. Constraints on this operator are given for interactions with electrons (purple) and quarks (blue), which come from the LEP and LHC collision experiments. The green band indicates couplings which are $95\%$ excluded by a Planck result of $g_* = 3.50 \pm 0.12$. The gray region for $\Lambda \gtrsim 5$ TeV corresponds to models which decouple during the QCD phase transition. The provided values of $\Delta g_*$ should therefore only be interpreted qualitatively in that region. The results for scalar, pseudoscalar and axial couplings are effectively the same. The results for a Dirac fermion are double those given in this figure, indicating that they must have decoupled during or before the QCD phase transition to be compatible with the Planck data.}
\label{fig:4fv-results}
\end{figure}

For each of these interactions, therefore, we can scan over possible effective suppression scales. The resulting contribution to $g_*$ as a function of $\Lambda$ is given in figure \ref{fig:4fv-results} for only the vector coupling, as the results for all four models are equivalent to within $5\%$. Therefore, any distinction between these models is below the experimental resolution of Planck. We also assume identical couplings to electrons and muons. For the possible case of flavor-specific couplings, the resulting $\Delta g_*$ will be the same as the flavor-blind case, where the equivalent flavor-blind $\Lambda$ is the smallest flavor-specific $\Lambda$.

The strongest experimental constraints on such couplings are indicated in figure \ref{fig:4fv-results} for both electron and quark interactions. These bounds come primarily from $\slashed{E}_T$ + monojet/monophoton searches at LEP and the LHC, again assuming universal coupling to quarks. The LHC bounds specifically came from $10$ fb$\phantom{}^{-1}$ of data, so we expect these experimental results to improve in the near future. Details of these exclusion limits can be found in \cite{Atlas5fb,Atlas10fb,FoxEtAl}. Couplings to muons are largely unconstrained in a flavor-specific model, but in the universal coupling case, constraints on any species would therefore limit the muon interactions.

As we see, theories with effective suppression scales $\Lambda \gtrsim 5$ TeV decouple prior to the muon entropy redistribution and are therefore predominantly affected by the QCD phase transition. As such, our results beyond those scales can only place an approximate upper bound on the possible contribution $\Delta g_*$. However, there is a range of potential suppression scales below 5 TeV but above the current experimental bound which is compatible with the constraints coming from a Planck measurement of $g_* = 3.50 \pm 0.12$. A model with a light Weyl fermion with dimension-6 interactions with SM fermions is therefore a viable model for substantial contributions to $g_*$. Our results indicate that a Dirac fermion contributes double what a Weyl fermion does at the same $\Lambda$, and that scalar, pseudoscalar, vector and axial vector operators give the same results to within 5\%. Consequently, Dirac fermions must have decoupled before or during the QCD phase transition in order to be compatible with the data from Planck. Future results from the LHC will continue to probe these interaction scales, providing an independent means of discovery or exclusion of such models.


\subsection{Spin-$1$: Gauge boson}
\label{sec:vector}


A massless spin-1 particle has fewer degrees of freedom than a massive one, and thus perturbative quantum effects cannot generate a mass, rendering a massless spin-1 particle technically natural. These gauge bosons are then automatic candidates for new light species. While gauge bosons can potentially acquire a mass through the Higgs mechanism, masses at scales $\lesssim$ eV are generically unnatural, unless there is a more complicated particle content\footnote{It is, of course, possible to Higgs the group at the TeV-scale, but have such a small gauge coupling that its mass is sub-eV ($g \lesssim 10^{-12}$). However, such a small gauge coupling implies that it will only recouple at very low temperatures, and even then, only to neutrinos. As neutrinos would have already decoupled from the SM, such interactions can only redistribute the neutrino energy density and cannot increase the total energy density. Thus there are no contributions of such a model to $g_*$.}. However, such non-minimal solutions are beyond the scope of this work, so we assume that any additional vector bosons are precisely massless. Similar to the case of a Goldstone boson, the corresponding gauge structure automatically restricts the available interactions for light spin-1 particles. The only possible operators are direct gauge couplings or dipole moment interactions with SM fermions, as well as kinetic mixing with SM gauge bosons.


\subsubsection{Kinetic mixing and gauge interactions}
\label{sec:u1prime}

As we will show, new massless gauge bosons with renormalizable couplings to SM fermions are viable candidates for contributions to $g_*$. Long-range force constraints greatly restrict the possible direct couplings of SM fermions charged under new gauge groups, such that these interactions must be too weak to contribute to $g_*$ \cite{BlinnikovEtAl}. However, such couplings can still arise due to kinetic mixing between the new and SM gauge fields. For such mixing to give rise to non-negligible $\Delta g_*$, there must also be new fermions charged under the new gauge group. The additional fermions obtain millicharged couplings to SM gauge fields, with astrophysical constraints such that these fermions must have masses $\gtrsim$ MeV. Such models are sensitive to the details of the full UV theory, as the hidden sector must come into equilibrium with the SM after originally being completely decoupled. The class of viable models is then constrained to a particular region of model-dependent parameter space.

For minimality, we consider the addition of a single new $U(1)$ gauge boson $A'$, with associated field strength $A'^{\mu \nu}$. The new field $A'$ can kinetically mix with the hypercharge gauge boson $B$ with the following operator

\begin{equation}
\mathcal{L} \supset -\frac{\epsilon}{2} A'^{\mu \nu} B_{\mu \nu} \textrm{,}
\label{eq:mixing}
\end{equation}

{\noindent}where $\epsilon$ is simply a dimensionless mixing parameter. Such hidden sector $U(1)$ gauge bosons which mix with hypercharge arise naturally in many models \cite{Holdom,Balasubramanian:2005zx,Abel:2008ai,GoodsellEtAl,Cicoli:2011yh,NelsonEtAl,KaplanEtAl,BelotskyEtAl, FrancaEtAl}.

This term indicates that our originally defined fields $A'$ and $B$ are not propagation eigenstates, and must be redefined to diagonalize the propagation basis. If both gauge bosons are precisely massless, then there is always a linear combination of gauge fields which does not couple to the SM. We can always define this linear combination as $A'$ and the orthogonal combination as $B$, such that $A'$ does not couple to the SM.

However, if $A'$ originally interacts with some new fermion $\chi$, any field redefinition will generically result in couplings between $\chi$ and the SM gauge bosons. The new fermion can then act as an intermediary between $A'$ and the SM, keeping all species in equilibrium. The most minimal theory involving new direct gauge couplings must contain both a new gauge boson $A'$ and a new Dirac fermion $\chi$, with the resulting interaction terms

\begin{equation}
\begin{split}
\mathcal{L} \supset & -\epsilon g_A \cos \theta_W \bar{\chi} \slashed{A} \chi - \epsilon g_A \sin \theta_W \bar{\chi} \slashed{Z} \chi - g_A \bar{\chi} \slashed{A'} \chi \textrm{.}
\end{split}
\label{eq:APlagrangian}
\end{equation}

We see that after the field redefinition $\chi$ interacts with both the photon and $Z$, with interaction strength that depends on the coupling $g_A$ of $\chi$ to $A'$, the original mixing $\epsilon$ between $A'$ and $B$, and the weak mixing angle $\theta_W$.

This particular choice of basis is technically arbitrary. It is also possible to redefine gauge fields such that the SM fermions possess millicharged couplings to $A'$ and $\chi$ possesses no couplings to the photon. The physics must be and is independent of the choice of basis. These rotations do not affect any physical observable, provided the observable is phrased in a basis-independent manner. Thermodynamic observables such as the overall energy density of massless gauge bosons, and therefore their contribution to $g_*$, are also basis-independent. We specifically choose to work in the basis of eq.\ (\ref{eq:APlagrangian}), where $A'$ does not interact with the SM, because the resulting early universe thermodynamics are more transparent. However, it is important to stress that the same results are true, but less obvious, in other bases.


The dominant interactions between $\chi$ and the SM are dimension-4 gauge couplings with the photon, as any interactions with the $Z$ are suppressed at temperatures below the weak scale. Dimensional analysis then implies that at temperatures large compared to $m_\chi$, the interaction rate is linear in temperature, $\Gamma_\chi \sim T$. Similar to the Goldstone couplings to leptons, this means that at high temperatures $\chi$ will be fully decoupled from the SM and then potentially recouples as the universe cools and the expansion rate drops when $\epsilon \ll 1$. Unlike the Goldstone case, $\chi$ is always interacting with $A'$, provided the $A'$ coupling is sufficiently large, such that $\chi$ and $A'$ can maintain equilibrium distributions. Therefore, the hidden sector has a well-defined temperature. The precise ratio of temperature of the hidden sector to the temperature of the SM prior to the recoupling of the two sectors is model-dependent, as more complicated hidden sectors will generally result in a wide range of possible temperatures. Consequently, we choose to explore a wide range of such initial ratios.

The thermodynamics are sensitive to whether $A'$ and $\chi$ are in equilibrium, rather than the precise coupling $g_A$, so we can simply fix the value of $g_A$ to be sufficiently large, without loss of generality. We select the value $g_A^2 = 0.1$, but our final results can be simply related to other values of $g_A$. Once $g_A$ is fixed, there are only three remaining parameters that can change: the kinetic mixing $\epsilon$, the new fermion mass $m_\chi$, and the ratio of initial temperatures $T_{hid}/T$.



Multiple star and supernova cooling observations, as well as various collider results, place significant constraints on millicharged fermions (for details see \cite{GoodsellEtAl,JaeckelEtAl}). Specifically, models with $m_\chi \lesssim 100$ keV are restricted to $\epsilon \lesssim 10^{-13}$, such that these species will never thermally couple to the SM for all reasonable initial values of $T_{hid}$. Light millicharged fermions can therefore not directly contribute to $g_*$, but more massive fermions can instead indirectly alter the CMB by maintaining equilibrium between the SM and $A'$, which then contributes a nonnegligible $\Delta g_*$. However, for this to occur, we need $m_\chi \lesssim 150$ MeV, such that $\chi$ is still present below the QCD phase transition. This therefore limits us to a very narrow range of allowed masses $m_\chi$ for models of millicharged species which affect the CMB. For models of this type, $\chi$ must couple to the SM prior to or during its annihilation, otherwise the hidden sector will again never couple to the SM. This limits the possible values for $\epsilon$ and $T_{hid}$ for any given mass $m_\chi$, in addition to constraints placed by independent observational and experimental bounds.

\begin{figure}[t]
\centering
\begin{tabular}{cc}
\includegraphics[width=8cm]{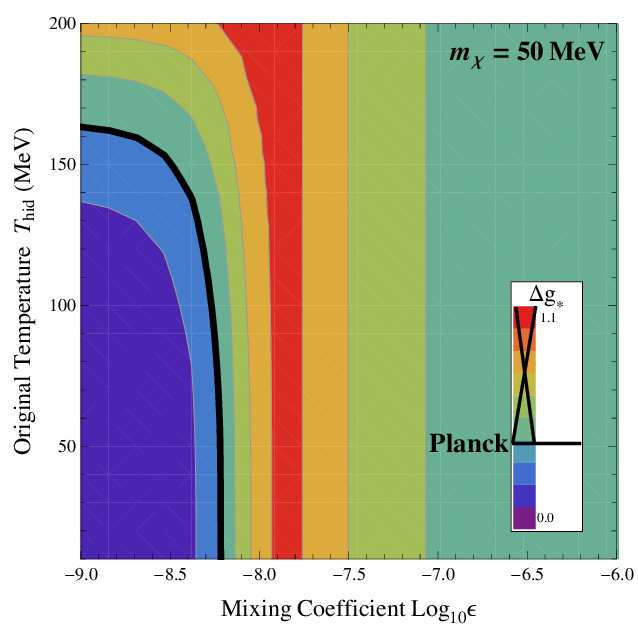}
&
\includegraphics[width=8cm]{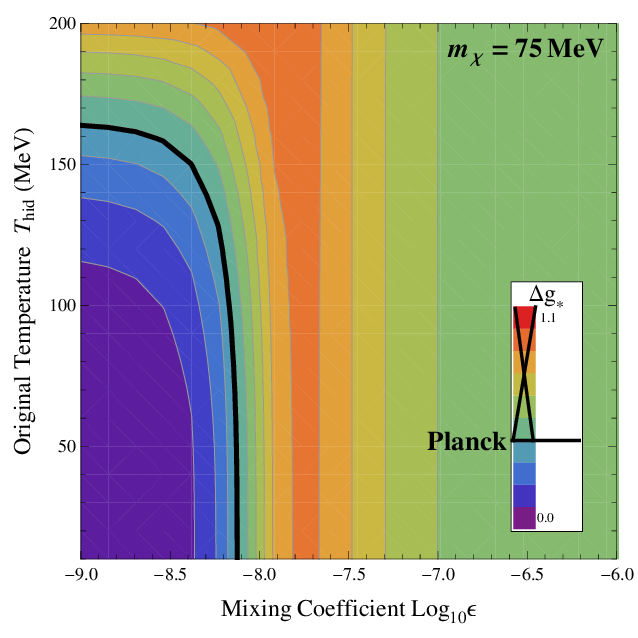}
\\
\includegraphics[width=8cm]{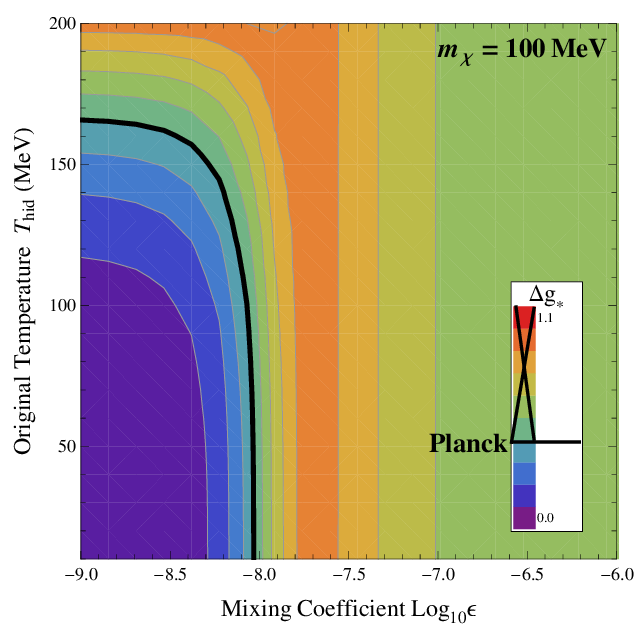}
&
\includegraphics[width=8cm]{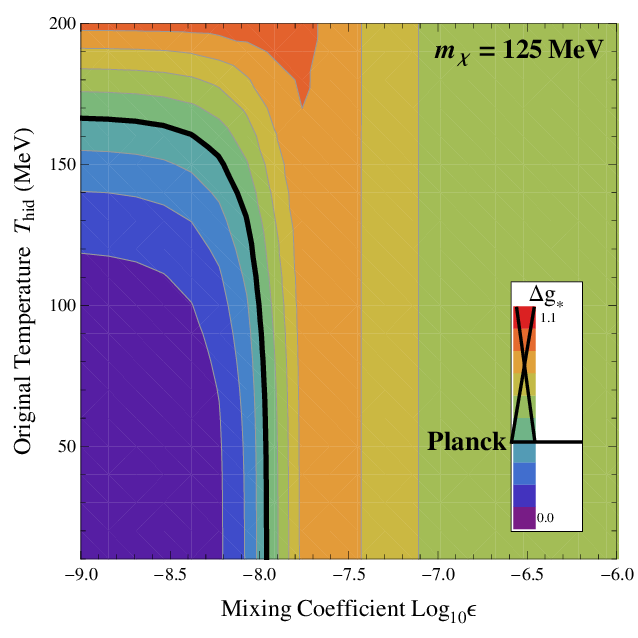}
\end{tabular}
\caption{$\Delta g_*$ due to a single gauge boson which couples to a new fermion with mixing-induced SM gauge couplings. The contribution to $g_*$ at recombination is given as a function of both the mixing parameter $\epsilon$ and the hidden sector temperature $T_{hid}$ when the SM temperature $T = 200$ MeV. Results are presented for (a) $m_\chi = 50$ MeV, (b) $m_\chi = 75$ MeV, (c) $m_\chi = 100$ MeV, and (d) $m_\chi = 125$ MeV. Blue and purple regions to the left of the black line are allowed by a Planck measurement of $g_* = 3.50 \pm 0.12$, although these regions were never in thermal equilibrium with the SM. Regions to the right of the black line are excluded by this result from Planck.}
\label{fig:charged}
\end{figure}

To illustrate the general behavior of these models, we consider four possible fermion masses within the allowed mass range. The corresponding results are shown in figure \ref{fig:charged}. In each case, the millicharged fermion has mass $m_\chi \gtrsim 10$ MeV, which are unconstrained by star and supernova cooling observations and therefore have the largest available ranges for $\epsilon$ and $T_{hid}$. The lowest mass shown in figure \ref{fig:charged} is actually $m_\chi = 50$ MeV, as the results are equivalent for masses between 10-50 MeV. For each of these cases, we scan over possible values for the mixing parameter $\epsilon$, as well as possible values for the original hidden sector temperature $T_{hid}$ when the SM temperature $T = 200$ MeV. We specifically consider $T_{hid}$ below the SM temperature $T$, assuming a minimal hidden sector model containing less particle content than the SM. The hidden sector will thus be colder due to fewer entropy redistributions. This procedure involved a modified version of the original code, the details of which can be found in appendix \ref{sec:code}.

As we see in figure \ref{fig:charged}, there is a basic pattern to the dependence of $\Delta g_*$ on both $\epsilon$ and $T_{hid}$. For very small values of $\epsilon$, the hidden sector is never coupled to the SM, and $A'$ receives all of the $\chi$ entropy redistribution. The contribution to $g_*$ is then dependent solely on the energy available in the hidden sector. The energy density increases as the initial temperature increases relative to the SM temperature. Initial temperatures of $T_{hid} \gtrsim \frac{4}{5} T$ are excluded by a Planck result of $g_* = 3.50 \pm 0.12$.

For increasing values of $\epsilon$, the hidden sector begins to couple with the SM, until at large values the two sectors quickly become fully coupled, regardless of the initial temperature $T_{hid}$. In this regime, the contribution of $A'$ to $g_*$ is precisely that of a new gauge boson which is originally coupled to the SM then decouples before the electron annihilation, $\Delta g_* \approx 0.5$.

In the transitional region from completely decoupled to completely coupled, the contribution rapidly climbs to $\Delta g_* \approx 1$, then rapidly decreases to the fully coupled limit for large $\epsilon$. This enhanced contribution to $g_*$ corresponds to a fortuitous combination of coupling and mass values, in which $A'$ participates in the muon entropy redistribution but is able to receive all of the $\chi$ entropy. This occurs because $\chi$ briefly couples to the SM, sharing the muon entropy, then quickly decouples as the muon and $\chi$ number densities begin to plummet, such that the SM receives none of the $\chi$ entropy. The result is a superheated population of $A'$ bosons, which contain a large fraction of the total energy density.

While the majority of this behavior has been largely $m_\chi$-independent, we do observe a slight decrease in the transitional region values of $\Delta g_*$ as the $\chi$ mass increases. Larger fermion masses result in the hidden sector decoupling earlier from the SM, and therefore receiving less of the muon entropy. Finally, there are no major distinctions between $m_{\chi} \sim 20$ MeV and $m_{\chi} \sim 50$ MeV, as these masses are proximate to neither the muon nor the electron mass.

For initial hidden sector temperatures below $\frac{T}{20}$, the behavior will be largely unchanged from the low-temperature results presented here. Theories with small mixing parameters will remain fully decoupled and contribute negligibly to $g_*$, while theories with larger $\epsilon$ values will rapidly reach equilibrium with the SM, such that their contribution $\Delta g_*$ is insensitive to the initial temperature.

We find that for any value of the initial temperature, $\epsilon$ is restricted to be $\lesssim 10^{-8}$ when there is a Dirac fermion $\chi$ with masses between $10-150$ MeV, forcing the SM and the hidden sector to never have been in thermal equilibrium. In addition, a scenario with $m_{\chi} \lesssim 10$ MeV is inconsistent with constraints from star and supernovae cooling, and one where $m_{\chi} \gtrsim 150$ MeV causes the $A'$s to decouple before or during the QCD phase transition. The result in the absence of $\chi$ is the same result as obtained by raising the $\chi$ mass and integrating it out of the theory; there exists a basis in which there are no couplings between the hidden sector and the SM, thus preventing the thermalization of $A'$. The scenario of new gauge bosons which mix with SM hypercharge is therefore further constrained by results from Planck.


\subsubsection{Dipole moments}

While it is always possible to eliminate any mixing-induced renormalizable couplings between SM fermions and a new unbroken gauge boson $A'_\mu$, there could generically still be higher-order nonrenormalizable couplings after integrating out $A'$-SM interaction mediators in the full theory. If the low-energy effective theory contains no light species charged under $U(1)_{A'}$, then the dominant interactions between $A'$ and the SM are of the form

\begin{equation}
\mathcal{L} \supset - \, \frac{1}{M^2} A_{\mu \nu}' \psi_R^c \sigma^{\mu \nu} h^\dagger \psi_L + h.c. \textrm{,}
\label{eq:DMprime}
\end{equation}

{\noindent}where $A'^{\mu \nu}$ is again the associated field-strength tensor and $M$ is the mass scale associated with the heavy species integrated out of the theory. After EWSB, the expansion of the Higgs field about its expectation value $v$ will lead to dipole moment interactions of the form

\begin{equation}
\mathcal{L} \supset - \, \frac{v}{M^2} A_{\mu \nu}' \psi_R^c \sigma^{\mu \nu} \psi_L + h.c. \rightarrow - \, \frac{1}{\Lambda} A_{\mu \nu}' \psi_R^c \sigma^{\mu \nu} \psi_L + h.c. \textrm{,}
\end{equation}

{\noindent}where we have now defined an effective dipole scale $\Lambda \equiv \frac{M^2}{v}$. Without knowledge of the full UV theory, it's possible for the resulting dipole interactions to have generic flavor structure, rather than be flavor-blind. We consider all such structure in this section.

The induced dipole couplings for pions must involve a composite pion operator which is antisymmetric in its two Lorentz indices. If the quark dipole moments are flavor-blind, such that the up and down quarks have the same couplings to $A'$, then all such antisymmetric operators vanish. If, instead, the dipole couplings are not flavor-blind, interactions between $A'$ and pions will potentially appear. However, there is no symmetry protecting against renormalization of such operators. We then expect these pion interactions to be strongly renormalized, thereby preventing us from making robust predictions about such contributions to $g_*$. We therefore focus solely on the dipole couplings of $A'$ to leptons.

With the interaction Lagrangian of the form

\begin{equation}
\mathcal{L} \supset - \frac{1}{\Lambda_f} \bar{\Psi_f} \sigma^{\mu \nu}\Psi_f A_{\mu \nu}' \textrm{,}
\end{equation}

{\noindent}where $\Psi_f$ can either be an elementary lepton or a composite nucleon, we obtain the bounds

\begin{equation}
\begin{split}
\Lambda_e &\gtrsim 2.0 \times 10^{10} \textrm{ GeV,} \\
\Lambda_{p,n} &\gtrsim 9.8 \times 10^{9} \textrm{ GeV.}
\end{split}
\end{equation}

These bounds again come from star and supernova cooling, and details can be found in \cite{Dobrescu,Hoffmann,Raffelt96}. There are also constraints on off-diagonal couplings $\Lambda_{\mu e}$ coming from $\mu \rightarrow e + \slashed{E}$ \cite{Dobrescu}, which limit such couplings to

\begin{equation}
\Lambda_{\mu e} \gtrsim 2.3 \times 10^9 \textrm{ GeV.}
\end{equation}

For direct muon constraints, we again calculate the approximate supernova cooling bounds

\begin{equation}
\Lambda_{\mu} \gtrsim 2.7 \times 10^6 \textrm{ GeV.}
\end{equation}

\begin{figure}[t]
\centering
\includegraphics[width=15cm]{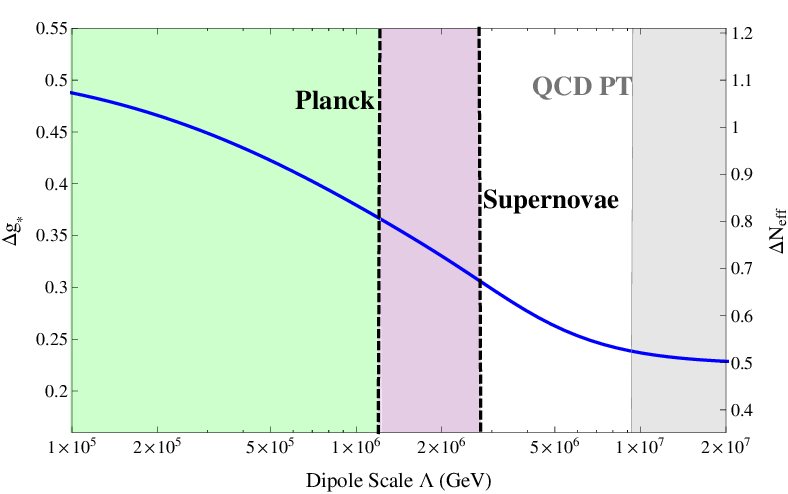}
\caption{$\Delta g_*$ due to an $A'$ which interacts primarily with muons. The contribution to $g_*$ at recombination is given as a function of the effective scale $\Lambda$, which suppresses this interaction. Constraints on this muon interaction resulting from observations of supernova cooling are given in purple, restricting $\Lambda \gtrsim 2.7 \times 10^6$. The gray region for $\Lambda \gtrsim 10^7$ GeV corresponds to models which decouple during the QCD phase transition. The provided values of $\Delta g_*$ should therefore only be interpreted qualitatively in that region. The green region corresponds to values of $\Lambda$ excluded by a Planck result of $g_* = 3.50 \pm 0.12$, which are comparable to, but slightly weaker than, the constraints placed by supernova cooling.}
\label{fig:aprimedipole-results}
\end{figure}

We find that the electron-only and electron-muon off-diagonal coupling scenarios are constrained to decouple before or during the QCD phase transition, preventing $A'$ from contributing to $g_*$ in this part of parameter space. However, when the coupling of $A'$ to those species in the SM present after the QCD phase transition is dominated by its coupling to the muon, there is still a potentially allowed range for $\Lambda_\mu$. Our results for muon-dominated couplings are shown in figure \ref{fig:aprimedipole-results}. We find that $\Lambda_\mu < 10^7$ GeV in order for the $A'$ to remain coupled after the QCD phase transition, and consequently contribute to $g_*$. This requires a significant hierarchy between the electron-$A'$ coupling and the muon-$A'$ coupling, but such a hierarchy is compatible with an MFV-like framework, as the hierarchy does not need to be much larger than $y_e/y_\mu$. Values of $\Lambda_\mu \lesssim 10^6$ GeV are inconsistent with a Planck result of $g_* = 3.50 \pm 0.12$, providing constraints which are approximately equivalent to those placed by supernova observations.

\subsection{Spin-$\frac{3}{2}$: Gravitino}

Any model of supergravity contains the gravitino, which is the unique elementary spin-$\frac{3}{2}$ particle. If supersymmetry were unbroken, the gravitino would be precisely massless. In a method similar to that of gauge symmetries, the spontaneous breaking of supersymmetry gives rise to a massless fermion, the Goldstino, which then becomes the longitudinal mode of the gravitino. As a result, the gravitino acquires a mass $m_{3/2} \sim \frac{F}{M_{pl}}$, where $F$ is generally the largest supersymmetry breaking scale squared in the theory. The gravitino can potentially remain a light degree of freedom for sufficiently low supersymmetry-breaking scales.

Na\"{i}vely, the gravitino would interact solely with gravitational strength and would therefore decouple at very high temperatures. However, at energy scales far above the gravitino mass, the Goldstino equivalence theorem ensures that the longitudinal components of the gravitino interact with Goldstino strength, potentially maintaining equilibrium with the SM down to lower temperatures. As is well known \cite{Wess:1992cp}, the Goldstino couplings to the SM are of the form

\begin{equation}
\mathcal{L} \supset - \frac{1}{F^2} \chi^{\dagger} \sigma_\mu \partial_\nu \chi T^{\mu\nu} \textrm{,}
\end{equation}

{\noindent}where $T^{\mu\nu}$ is the stress-energy tensor comprised of SM fields. While this coupling is no longer gravitationally suppressed, it is still a dimension-8 operator, such that the gravitino will still decouple above the QCD phase transition for all viable supersymmetry-breaking parameters $F$ and not contribute significantly to $g_*$.

\subsection{Spin-$2$: Graviton}

The unique elementary spin-$2$ particle is the graviton. The graviton interacts solely with gravitational strength, such that it either decouples from the SM at very high temperatures or is never even in thermal equilibrium. Similar to the discussion of subsection \ref{sec:relevanttemps}, the contribution to $g_*$ of gravitons which decouple at such large temperatures is well below the sensitivity of Planck.


\subsection{Models with light masses}
\label{sec:massive}

\begin{figure}[t]
\centering
\begin{tabular}{cc}
\includegraphics[width=10cm]{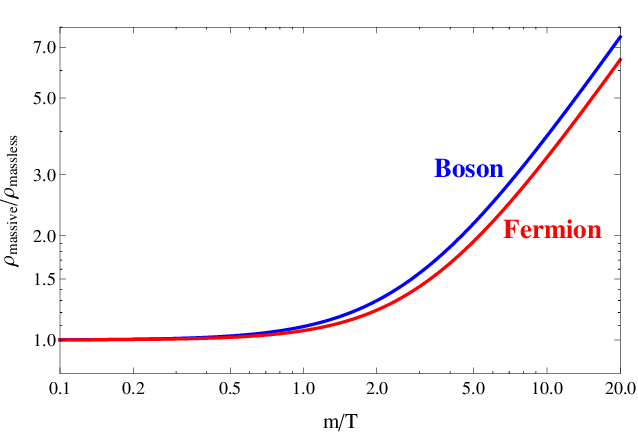}
\\
\includegraphics[width=10cm]{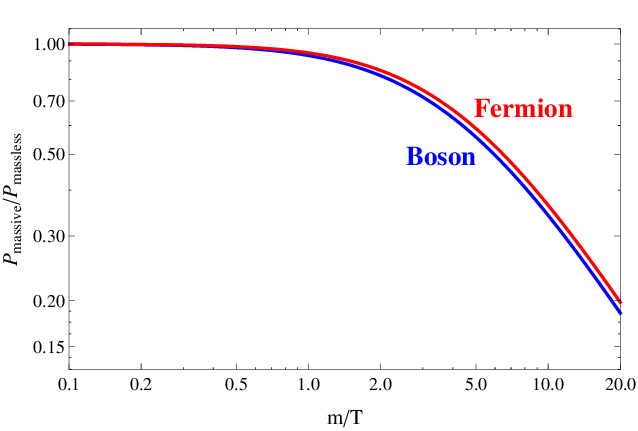}
\end{tabular}
\caption{Ratios of the (a) energy density $\rho$ and (b) pressure $P$ of a massive particle which decouples at some high temperature to those of a massless particle which decouples at the same temperature, expressed as a function of $\frac{m}{T}$, where $m$ is the particle mass and $T$ is the temperature of interest. These calculations assume that the particle decoupled such that it maintained an equilibrium distribution, specifically the Bose-Einstein (blue) or Fermi-Dirac (red) distribution. At temperatures below the mass of the particle, the pressure of the massive particle rapidly drops, while the energy density rapidly becomes much larger than that of a massless particle. The resulting deviations of physical observables, such as the expansion rate $H$, can be extracted from these ratios to see the sensitivity of such observables to the particle's mass.}
\label{fig:massive}
\end{figure}

Up to this point, we have considered any new species to be precisely massless, which allows their contribution to $g_*$ to be directly computed from the distribution function near recombination. This approximation is valid for any particles with masses $m \ll$ eV, for these particles will still be fully relativistic during and shortly after the formation of the CMB. This range of validity can be explicitly seen in figure \ref{fig:massive}, which shows the ratios of both the energy density and pressure of a massive particle which decouples at high temperatures to those of a massless particle which decouples at the same high temperature, all as a function of the particle's mass over the relevant temperature. For any given temperature, such as that of recombination, we can then use these simple ratios to determine the range of masses which can be treated as negligible, such that our massless approximation is valid.

It is still possible for there to be natural models with $m \sim$ eV. One example is the addition of 1-3 light sterile neutrinos, which are motivated by multiple short baseline oscillation results suggesting the existence of neutrino mass splittings distinct from those required to fit solar and atmospheric neutrino data (for details, see \cite{AbazajianEtAl} and references therein). Various analyses of these sterile neutrino models can be found in \cite{MohapatraEtAl,Nelson, KuflikEtAl,HoEtAl,CirelliEtAl,KraussEtAl,HannestadEtAl,MirizziEtAl,Jacques:2013xr}.

The presence of nonzero masses alters the relation between the energy density and the pressure of a species, such that the full effects cannot be captured by a single number $\Delta g_*$. In order to interpret the constraints of CMB measurements on species which become nonrelativistic during recombination, we must consider the resulting differences between Silk damping and early ISW. Both of these processes are sensitive to the precise evolution of $H$, whose time-dependence is sensitive to the mass of new light species. In addition, there are new mass-dependent effects which can arise, such as alterations to the matter power spectrum and to gravitational lensing of the CMB, which are similar to the effects caused by nonzero neutrino masses. Such discussion is beyond the scope of our current work, but more details can be found in \cite{Planck,Lesgourgues12}.

Silk damping is primarily sensitive to the overall expansion rate, and therefore the overall energy density, near the point of recombination. Any additional light species will add more energy density than is predicted solely by the SM. However, if such species have non-negligible masses, these new particles behave as relatively hot dark matter, as they will have become nonrelativistic by the modern era. Consequently, they contribute to measurements of $\Omega_{DM}h^2$ {\it today}, whereas they did not impact the CMB in the same fashion as standard cold dark matter. The exact contribution of a massive species to Silk damping is therefore sensitive to the amount of dark matter in our universe, which is dominated by uncertainty in the overall dark matter content, and a more careful analysis of the effects of new light species is needed.

Similarly, the early ISW effect is sensitive to the radiation/matter ratio following the formation of the CMB. New massive species will be transitioning to a nonrelativistic distribution during this period, behaving as neither pure radiation nor pure matter. Again, the exact prediction of early ISW effects is also dependent on the precise energy density of cold dark matter.

The main complication to the calculation of $\Delta g_*$ for such models arises from the use of the specific $\Lambda$CDM framework in calculating cosmological parameters from CMB data, in which the mass of dark matter is significantly higher than the temperature of recombination. This leads to model-dependence in the reported bounds, which do not necessarily exclude models which fall outside of this framework.

It is important to stress that the difficulty arises due to uncertainty in the precise expansion rate and dark matter content, not due to any calculational uncertainty in the new light species sector. For example, one can precisely calculate the decoupling of light sterile neutrinos which potentially accomodate the recent short baseline results. We used both the normal and inverted hierarchy best-fit models of \cite{BlennowFernandez}, which includes two light sterile neutrinos, and calculated the evolution of the two mostly-sterile mass eigenstates. We find that, for both hierarchies, these species decouple from the SM near the end of the muon entropy redistribution, such that they would contribute $\Delta N_{eff} \approx 1.9$ if they were massless. However, due to their non-negligible mass, the effects of these particles on the CMB is not fully characterized simply by a contribution to $N_{eff}$ or $g_*$. In order to fully probe the effect of models such as this on the CMB, a more general analysis of the CMB anisotropy data must be taken, which includes the possibility of nonzero masses for various additional light species. Such an analysis is beyond the scope of this paper, but will be pursued in future work.


\subsection{Models without new light species}

Up to this point, we have considered the addition of new light species to the SM in order to increase the total relativistic energy density, $\rho_{rel}$, at recombination. The other possibility is a modification of the distribution functions of light species already present in the SM. The distribution function of photons is well-established as Bose-Einstein by measurements of the CMB, such that the energy density of photons at recombination is known to high precision. The only remaining option is therefore a modification of the distribution function of neutrinos. If this change were to occur while neutrinos were still in equilibrium with the SM, then interactions with SM species would thermalize the distribution functions, washing out any original alteration. Therefore, new physics must only affect the neutrino energy density at temperatures below the MeV scale. Here we briefly discuss the possible mechanisms which can alter the distribution function of neutrinos to increase $g_*$ at recombination: a neutrino asymmetry, interactions with new massive species, and new interactions between neutrinos and the remaining SM species.

A simple modification to the distribution function of a species is the introduction of a chemical potential $\mu$, such that

\begin{equation}
f(t,E) = \frac{1}{e^{(E - \mu)/T} + 1} \textrm{,}
\end{equation}

{\noindent}for fermions, with $\mu \rightarrow - \mu$ for antifermions. Such a chemical potential results in an asymmetry between the number of particles and antiparticles. Once a species has fully decoupled and freely evolves, the Boltzmann equation constrains $f$ to remain solely a function of $a(t)p$, such that $\xi \equiv \frac{\mu}{T}$ is then time-independent. We can then express any resulting effects in terms of this constant $\xi$.

Although the neutrino distribution function is no longer Fermi-Dirac after decoupling from the SM, we shall assume it is for illustrative purposes. The total energy density stored in neutrinos and antineutrinos with nonzero $\xi$ is given by

\begin{equation}
\rho_\nu = -\frac{3 N_\nu T_{\nu}^4}{\pi^2} \left(\mathrm{Li}_4(-e^\xi) + \mathrm{Li}_4(-e^{-\xi}) \right) = \frac{7 N_{\nu} \pi^2 T_\nu^4}{120} \left(1 + \frac{30\xi^2}{7\pi^2} + \frac{15 \xi^4}{7\pi^4}\right) \textrm{.}
\end{equation}

The presence of a nontrivial chemical potential for neutrinos would therefore increase the energy density, thereby increasing $g_*$. The electron neutrino chemical potential affects the neutron-to-proton ratio prior to the start of BBN through reactions of the form $p + \bar{\nu} \rightarrow n + \bar{e}$. This ratio directly affects the helium-4 abundance after BBN, and so bounds can be placed on the electron neutrino chemical potential. Furthermore, since all neutrino mass eigenstates contain some wavefunction overlap with the electron neutrino, all of the neutrino mass eigenstate chemical potentials are constrained. The result is $\xi \lesssim 0.1$ for each of the three neutrino species (\cite{BargerEtAl,CastorinaEtAl} and references therein).


A second possibility is the interaction of some new massive species with neutrinos. This heavy species can alter the neutrino distribution through annihilation or decay \cite{Boehm:2012gr}. For the case of annihilation, the new species must interact predominantly with neutrinos and possess a mass $\lesssim 10$ MeV, such that the resulting entropy redistribution occurs after neutrinos decouple from the SM. For the case of decay, the heavy species must have fully decoupled at some higher temperature, leaving a significant relic energy density, with a decay rate such that it decays predominantly to neutrinos after neutrino decoupling but prior to recombination. These decays would then significantly alter the neutrino distribution, creating a large number of neutrinos with energies comparable to the particle mass. For both cases, the mass, number density, and coupling to neutrinos for this new species determine the precise contribution to $g_*$, making these scenarios highly model-dependent.


Finally, the existence of higher-dimensional operators coupling the neutrinos to other SM species could potentially maintain thermal equilibrium between neutrinos and the SM until lower temperatures. A later point of neutrino decoupling would result in a larger share of the electron entropy being distributed to neutrinos, raising their energy density. The possible interactions with the lowest dimensionality are electromagnetic dipole moments or four-fermion interactions between neutrinos and electrons, which are significantly constrained by star cooling \cite{Raffelt12} and the LEP collider \cite{FoxEtAl}.


\section{Conclusions}
\label{sec:conclusions}

The Standard Model of particle physics represents our current knowledge of the quantum field theory that best describes all short-distance interactions down to $10^{-17}$ cm.  Knowing that this model is incomplete leads us to search for fundamental particles outside the Standard Model. While the search for heavier particles continues at colliders, we focus on another class of new physics -- light, stable particles -- which can be probed via their effects on cosmology, most strikingly on the Cosmic Microwave Background.  In this article, we have surveyed what we call the most `natural' (or least contrived) models and their parameter spaces.  By doing so we lay out the reach of current and future experiments detailing the power spectra in the Cosmic Microwave Background and other probes of the initial density perturbations and cosmological parameters.

We have been able to analyze the effects on the radiation density of the universe of new light degrees of freedom which decouple after the QCD phase transition.  This includes species that decouple at `complicated' cosmological times, such as the time around which the muon becomes non-relativistic.  We are able to compute the energy density, and consequently $\Delta g_*$, to an accuracy of $1\%$. This allows us to place constraints on the couplings in those well-motivated BSM effective models which contain new light degrees of freedom, which are competitive with constraints coming from other areas of physics. We do this using a program which solves the Boltzmann and Friedmann equations for the case of one new light species, calculating the resulting evolution of that species' distribution function, while approximating the SM species using fully thermalized equilibrium distributions and only considering the effects of leading order interaction terms. Using these calculations, we have demonstrated the ability of Planck and future experiments to place exclusion limits on all natural, minimal models with new light species. The compatibility of each model with the recent Planck results is given in table \ref{tab:results}.

Higher levels of calculational accuracy could be achieved if we used a different numerical algorithm which was better adapted for the integro-differential equations considered in this paper, or used a larger and finer momentum grid. In addition, loop corrections to the amplitudes, three-body final states, and finite-temperature QFT effects all contribute at the $0.1\%$ level. If much higher precision is ever achieved observationally, potentially through next-generation polarization measurements, then these improvements would be warranted. Such a high-precision measurement of $g_*$ would better reveal degrees of freedom which decouple before or during the QCD phase transition. In such a scenario, this measurement, combined with independent measurements of the nature and couplings of a new light degree of freedom could potentially even allow us, in this way, to probe the structure of the QCD phase transition.

\begin{table}[t]
\centering
\begin{tabular}{|c|c|c|}
\hline
{\bf Model} & {\bf Operator} & {\bf Results} \\ \hline
Goldstone bosons & $\frac{1}{\Lambda} \partial_\mu \phi \bar{\Psi}\gamma^\mu\gamma^5 \Psi$ &
Flavor-blind: Decouple during/before \\
& & QCD PT \\
& & Muon-only: $\Lambda > 2 \times 10^3$ TeV \\ \hline
Four-fermion V & $\frac{1}{\Lambda^2}\chi^\dagger \bar{\sigma}^\mu \chi \bar{\Psi}\gamma_{\mu} \Psi$ & Weyl: $\Lambda > 1$ TeV \\
(S, P, A same to & $\frac{1}{\Lambda^2}\bar{\mathrm{\bf X}} \gamma^\mu \mathrm{\bf X} \bar{\Psi}\gamma_{\mu} \Psi$ & Dirac: $\Lambda > 5$ TeV \\ 5\%; see text) & & \\ \hline
$U(1)'$ & $\epsilon e \bar{\chi}\slashed{A}\chi$ & $\epsilon < 10^{-8}$ for $10$ MeV $\leq m_\chi \leq 150$ MeV \\ & & $m_\chi > 150$ MeV: Decouple during/before \\ & & QCD PT \\ \hline
$A'$-dipole & $\frac{1}{\Lambda} A'_{\mu\nu} \bar{\Psi}\sigma^{\mu\nu}\Psi$ & Flavor-blind: Decouple during/before \\
& & QCD PT \\
& & Muon-only: $\Lambda > 3 \times 10^3$ TeV \\ \hline
Massive Particles & Any & Inconclusive; mass-dependent \\
(e.g. Sterile Neutrinos) & & \\ \hline
\end{tabular}
\caption{Compatibility of those natural, minimal models considered here with the recent results of the Planck satellite, $g_* = 3.50 \pm 0.12$ and $N_{eff} = 3.30 \pm 0.27$ \cite{Planck}. While the current Planck results are in tension with other observational measurements, future experiments will greatly improve the precision and reach of these exclusion limits.}
\label{tab:results}
\end{table}

The future work we intend to pursue is the inclusion of the mass effects on different observables in the Cosmic Microwave Background.  While this is only relevant in a narrow mass range (close to recombination temperatures), it turns out to be quite important for a number of specific models, such as those of sterile neutrinos. The more accurately we can describe their impact on the ISW effect and on Silk damping, the greater the possibility of finding a `smoking gun' for such models.

If we coarsely divide the types of possible undiscovered particles into four types, categorized by stable or unstable and light or heavy, this work is an attempt to help push forward our probe of one of these types -- new stable light particles.  As the challenge to build new, more powerful high-energy colliders intensifies, it is exciting to see this new frontier mature as an additional source of information about the world beyond the Standard Model.

\appendix

\section{Notations and conventions}

We take the metric signature to be $(+,-,-,-)$. We set $\hbar = c = k_B = 1$ and give temperatures in units of energy. We use $f_{\pi} \approx 93$ MeV. We use $M_{pl} \approx 1.22 \times 10^{19}$ GeV, with $G = M_{pl}^{-2}$. $a(t)$ represents the scale factor of the universe. We use both Weyl and Dirac notation for fermions, depending on need. Where there is an ambiguity, we use capital $\Psi$ and $\mathrm{\bf X}$ for Dirac spinors, and lowercase Greek $\psi$s and $\chi$s with subscripts $_L$ and $_R^c$ for Weyl spinors. All of our Weyl spinors are left-handed. We use the notation of (the mostly minus version of) \cite{Dreiner:2008tw} for two-component spinor Feynman rules. The SM Higgs doublet is called $h$, and SM fermions are generically referred to as $\psi$, when not referencing a specific fermion. All BSM light fields not coming from special UV completions are given by $\phi$ for scalars, $\chi$ for fermions and $A'$ for gauge bosons, with associated field strength $A'^{\mu \nu}$.

\section{Details of code}
\label{sec:code}

In order to solve the Boltzmann equation to obtain the distribution functions and consequently $\Delta g_*$, a numerical code was written in C++. The code evolves a universe forward in time subject to some initial boundary conditions. The inputs to the code are the masses $m_i$, statistics $\sigma_i$ and degrees of freedom of species $g_i$ in the universe, as well as the initial temperature $T_i$. Interactions between the various species in the SM sector are strong enough that it is safe to assume that the distribution functions are Fermi-Dirac or Bose-Einstein until down to temperatures well below their mass, as discussed in subsection \ref{sec:thermo}. At that point, their number density has become low enough that their interactions to our new species have frozen out, and we work in an effectively radiation-dominated universe, so the error in the distribution function does not affect our evolution. In order to work with $\mathcal{O}(1)$ numbers for the distribution functions which are decoupling, we track $v(p, t)\equiv v$ instead, defined implicitly through the equation $f(p, t) = \frac{1}{e^{v} - \sigma}$, where $\sigma$ is $1$ for bosons and $-1$ for fermions. The code solves for the following quantities:

\begin{itemize}
\item $v_i(p, t)$ for all BSM species $i$
\item $T_{SM}=T_\gamma \equiv T$ (as we work above the neutrino decoupling temperature)
\item $\rho_i$, $P_i$, $n_i$ for all species $i$
\item $H$ and $a$
\end{itemize}

The following equations are used to solve for the aforementioned quantities:

\begin{equation}
E \frac{\partial v}{\partial t} - H p E\frac{\partial v}{\partial p} = \frac{\partial v}{\partial f} C[f] \textrm{,}
\label{eqn:boltzmann}
\end{equation}

{\noindent}where we discuss computation of $C[f]$ below,

\begin{equation}
\label{eqn:T}\frac{\partial \rho_{tot}}{\partial t} = y_{SM} \frac{\partial T_{SM}}{\partial t} + \frac{\partial \rho_\chi}{\partial t} = -3H (\rho_{tot} + p_{tot}) \textrm{,}
\end{equation}

{\noindent}where $y_{SM} = \sum_{i\subset SM} y_i$ and

\begin{equation}\label{eqn:y}y_i = \frac{g_i}{2\pi^2}\frac{1}{T^2} \int_0^\infty dp~ p^2 E^2 e^{v_i} f_i^2\textrm{,}\end{equation}

\begin{equation}\label{eqn:H}H = \sqrt{\frac{8\pi G}{3} \rho_{tot}}\textrm{,}\end{equation}

\begin{equation}\label{eqn:a}\frac{\partial a}{\partial t} = aH\textrm{,}\end{equation}

\begin{equation}\rho_i = \frac{g_i}{2\pi^2} \int_0^\infty dp ~ p^2 E f_i\textrm{,}\end{equation}
\begin{equation}P_i = \frac{g_i}{2\pi^2} \int_0^\infty dp ~ \frac{p^4}{3E} f_i\textrm{,}\end{equation}
\begin{equation}n_i = \frac{g_i}{2\pi^2} \int_0^\infty dp ~ p^2 f_i\textrm{,}\end{equation}
\begin{equation}g_{*,i} = \frac{30\rho_i}{\pi^2 T^4}\textrm{.}\end{equation}

The various quantities are tracked over 2000 timesteps, spaced logarithmically. We begin at $T = 200$ MeV and typically end at around $1$ MeV. As we only have earlier time information, time derivatives that cannot be computed analytically are typically computed by forming an interpolating polynomial to the previous four pieces of data and taking an exact derivative of the resulting polynomial. Furthermore, this technique is also used to obtain an estimate of the next value of the variable in question, in order to improve the accuracy of the code. We found that this technique is considerably more accurate at numerically evaluating derivatives than more elementary finite difference methods.

Our distribution function is evaluated on a grid of $100$ momentum-steps, logarithmically spaced between $10$ keV and $10$ GeV. All derivatives with respect to momentum that cannot be computed analytically are computed with the interpolating polynomials method described previously. 

The algorithm used is as follows:

\begin{itemize}
\item Set boundary conditions: Bose-Einstein or Fermi-Dirac distributions for every species with temperature $T_i$, and $a_i = 1$
\item Compute all initial $\rho_i$, $p_i$, $n_i$, $H$, $g_{*,i}$
\item Compute the next SM temperature with eq.\ (\ref{eqn:T})
\end{itemize}

Iterate the following at timestep $j$ until 2000 timesteps:

\begin{itemize}
\item Approximate $H_j$ at the next timestep by extrapolating the previous values of $H(t)$
\item Solve the Boltzmann equations of all BSM species $i$ for $v_{i,j}$, described more thoroughly below
\item Compute all remaining undetermined parameters, as well as $H(t_j)$
\item Compute the next SM temperature with eq.\ (\ref{eqn:T})
\end{itemize}

The Boltzmann eq.\ (\ref{eqn:boltzmann}) was solved using a generalization of a predictor-corrector method. The objective was to simultaneously vary $v_{i}$ at all points on the momentum grid, attempting to minimize the quantity

\begin{equation}\sum_{i\subset \mathrm{BSM}}\sum_{k = 1}^{100} \frac{1}{p_k} \left|E_k \frac{\partial v_i(p_k)}{\partial t} - H p_k E_k\frac{\partial v_i(p_k)}{\partial p_k} - \frac{\partial v_i}{\partial f_i}|_{p_k} C(f_i(p_k))\right|\textrm{.}\end{equation}

Because the collisional integral is the most computationally intensive part of the algorithm, we attempted to minimize the number of calls of it. This was accomplished by primarily studying the effects of the variation of $v_i$ on the left-hand side of the Boltzmann equation, as the right-hand side varied more slowly, only recomputing $C$ when we had settled on a $v$ that minimized the local error

\begin{equation}\left| E_k \frac{\partial v_i(p_k)}{\partial t} - H p_k E_k\frac{\partial v_i(p_k)}{\partial p_k} - \frac{\partial v_i}{\partial f_i}|_{p_k} C(f_i(p_k)) \right|\textrm{,}\end{equation}

{\noindent}with a relative accuracy of $\approx 10^{-6}$.

The collisional integral in eq.\ (\ref{eqn:collisionalint}) has been computed by following the method devised by \cite{HannestadMadsen}. We briefly summarize the algorithm; see \cite{HannestadMadsen} for more details. Define the following quantities:

\begin{itemize}
\item The angle between $\vec{p}_1$ and $\vec{p}_2$ is $\alpha$
\item The angle between $\vec{p}_1$ and $\vec{p}_3$ is $\theta$
\item The azimuthal angle between $\vec{p}_2$ and $\vec{p}_3$ is $\beta$
\item $x = \cos\alpha$
\item $z = \cos\theta$
\item $Q = m_1^2 + m_2^2 + m_3^2 - m_4^2$
\end{itemize}

In the amplitude $|\mathcal{M}|^2$, we plug in

\begin{equation} p_1\cdot p_2 = E_1 E_2 - |\vec{p}_1| |\vec{p}_2| x\textrm{,}\end{equation}
\begin{equation} p_1\cdot p_3 = E_1 E_3 - |\vec{p}_1| |\vec{p}_3| z\textrm{,}\end{equation}
\begin{equation} p_1\cdot p_4 = m_1^2 + E_1 E_2 - |\vec{p_1}| |\vec{p}_2| x - E_1 E_3 + |\vec{p}_1| |\vec{p}_3| z\textrm{,}\end{equation}
\begin{equation} p_2\cdot p_3 = E_1 E_2 - |\vec{p}_1| |\vec{p}_2| x - E_1 E_3 + |\vec{p}_1| |\vec{p}_3| z + Q/2\textrm{,}\end{equation}
\begin{equation} p_2\cdot p_4 = E_1 E_3 - |\vec{p}_1| |\vec{p}_3| z + m_2^2 - Q/2\textrm{,}\end{equation}
\begin{equation} p_3\cdot p_4 = E_1 E_2 - |\vec{p}_1| |\vec{p}_2| x - m_3^2 + Q/2\textrm{.}\end{equation}

We can change variables to $|\vec{p}_1|$, $|\vec{p}_2|$, $|\vec{p}_3|$, $\vec{p_4}$, $x$, $z$, $\beta$ and $\mu$, where $\mu$ is an integration variable parameterizing the $SO(2)$ rotational symmetry about $\vec{p}_1$. The collisional integral has no dependence on $\mu$, and it can therefore be done trivially. After using the momentum-conserving delta function to integrate $\vec{p}_4$, we can use the energy-conserving delta function to integrate $\beta$. Now that there are no more four-vectors in our expression, we switch notation $|\vec{p}_i| \rightarrow p_i$. After some algebra, it has been shown that $C$ can be written in the form

\begin{equation}C(f(p_1)) = \int_0^\infty dp_2 \int_0^\infty dp_3 \frac{p_2^2 p_3^2 \Omega({f}) F}{(2\pi)^5 16 E_2 E_3} \textrm{,} \end{equation}

{\noindent}where $\Omega$ was defined before as $f_3 f_4 (1 + \sigma_1 f_1)(1 + \sigma_2 f_2) - f_1 f_2 (1 + \sigma_3 f_3) (1 + \sigma_4 f_4)$ and $F = F(p_1,p_2,p_3)$ is

\begin{equation}F = \int_{-1}^1 dz \int_{x_-}^{x_+} dx \frac{|M|^2(x,z)}{\sqrt{a(z) x^2 + b(z) x + c(z)}}\Theta(A)\textrm{,}\end{equation}

{\noindent}where

\begin{equation}a(z) = \left(-4p_2^2(p_1^2+p_3^2)\right) + \left(8 p_2^2 p_1 p_3\right) z \textrm{,}\end{equation}
\begin{equation}b(z) = \left(p_1 p_2 (8 \gamma + 4 Q)\right) + p_2 p_3 \left( 8 p_1^2 - 8\gamma - 4Q\right) z + \left(-8p_1 p_2 p_3^2\right) z^2\textrm{,}\end{equation}
\begin{equation}c(z) = \left(4 p_2^2 p_3^2 - 4\gamma^2 - 4\gamma Q - Q^2\right) + \left(-p_1 p_3(8\gamma  + 4 Q)\right) z + \left( -4 p_3^2( p_1^2 + p_2^2)\right) z^2\textrm{,}\end{equation}
\begin{equation}\gamma = E_1 E_2 - E_3(E_1 + E_2)\textrm{,}\end{equation}
\begin{equation}x_\pm = \frac{-b \mp \sqrt{b^2 - 4ac}}{2a}\textrm{.}\end{equation}

Note that since $a \leq 0$, we know that $x_+ \geq x_-$. If we define 

\begin{equation}z_\pm = \frac{1}{2p_1 p_3} \left(-2\gamma - 2p_2^2 -Q \pm 2p_2 \sqrt{2\gamma + p_1^2 + p_2^2 + p_3^2 + Q}\right)\textrm{,}\end{equation}

then $\Theta(A)$ is $1$ when $2 \gamma + p_1^2 + p_2^2 + p_3^2 + Q > 0$, $z_+ > -1$, and $z_- < 1$, and is $0$ otherwise.

The amplitudes $|\mathcal{M}|^2$ were computed with the assistance of Tracer \cite{JaminEtAl}. After the substitutions above, $|\mathcal{M}|^2$ can be written as a rational function in $x$ and $z$. We first expand it in $x$, and integrate it analytically with Mathematica. Afterwards, if it is possible to analytically integrate with respect to $z$, we do so and store the results at all $10^6$ combinations $\{p_1,p_2,p_3\}$. Otherwise, we numerically integrate with respect to $z$ at all $10^6$ points in the resulting phase space. These are stored and then loaded into our C++ code.

The code has been verified to give the same answer for $T_{eff,\nu}(p)$ as that given in \cite{GnedinEtAl}, giving the same values for $\Delta g_*$ to the percent level. In addition, the results were computed and compared to all cases where it is possible to use the instantaneous decoupling approximation or otherwise solve the problem analytically, and agreement was again found to the percent level or better in all cases. Percent-level accuracy is more precise than the resolution of the Planck satellite, and so we do not quote theoretical errors throughout the paper.

In subsection \ref{sec:u1prime}, we reference a modified version of the code suitable for tracking two separate thermalized sectors which undergo partially thermalizing interactions. The structure of the code is very similar to the code outlined above, but with a few changes:

\begin{itemize}
\item The initial conditions are $T_{SM}$, $T_{hid}$ and $\epsilon$.
\item The distribution functions are always Bose-Einstein or Fermi-Dirac, and so only the two temperatures are tracked, eliminating the need for storing any distribution functions.
\item We use a modified set of equations to track the distribution functions:
\end{itemize}

\begin{equation}\frac{\partial T_{SM}}{\partial t} = \frac{-3H(\rho_{SM}+P_{SM}) - \Gamma}{y_{SM}} \textrm{,} \end{equation}

\begin{equation}\frac{\partial T_{hid}}{\partial t} = \frac{-3H(\rho_{hid}+P_{hid}) + \Gamma}{y_{hid}}\textrm{,}\end{equation}

\begin{equation}\Gamma = \frac{g}{2\pi^2} \int_0^\infty dp_1~ p_1^2~ C[f_1(p_1)]\textrm{.}\end{equation}

\begin{itemize}
\item A semi-implicit Euler method was used to compute the temperatures at the next timesteps, in order to minimize the amount of time spent computing $\Gamma$.
\end{itemize}

\acknowledgments{We are grateful to Ian Anderson, Masha Baryakhtar, George Bruhn, Fabrizio Caola, Cathy Caviglia, David Ely, Nick Gnedin, Oleg Gnedin, Chiu Man Ho, Marc Kamionkowski, Tobias Marriage, Colin Norman, Josef Pradler, Adam Riess, Dan Stolarski, Raman Sundrum, and Yuan Wan for helpful discussions. C.B., M.W., and D.K. were supported under grant PHY-1214000. C.B. was also partially supported under grant PHY-0968854. We are thankful to the University of Maryland, College Park Physics Department for computing time on the UMD HEP T3 Computing Cluster. C.B. is also grateful for support by the Maryland Center of Fundamental Physics.}

\bibliography{RDOF_Reformatted_References}
\bibliographystyle{JHEP}

\end{document}